\documentclass[12pt]{elsarticle}

\usepackage{graphicx}
\usepackage{amssymb}
\usepackage{color}
\usepackage{amsmath}

\def\vec{\boldsymbol}
\newcommand{\dive}{\vec \nabla \cdot}

\journal{International Journal of Heat and Fluid Flow}

\begin{document}

\begin{frontmatter}



\title
{Fractal scaling of turbulent premixed flame fronts: application to LES}


\author[dima]{F. Battista}

\author[enea]{G. Troiani}

\author[kth]{F. Picano\corref{cor1}}
\ead{picano@mech.kth.se}
\cortext[cor1]{Phone: +46 723 223 075. Fax: +46 8 205131}

\address[dima]{Department of Mechanical and Aerospace Engineering, \\ Sapienza University, Via Eudossiana 18, 
00184,  Rome, Italy.}
\address[enea]{Sustainable Combustion Laboratory, ENEA C.R. Casaccia, 
            Rome, Italy}
\address[kth]{KTH Mechanics, Royal Institute of Technology,\\ Osquars Backe 18, 100-44 Stockholm, Sweden}

\begin{abstract}
The fractal scaling properties of turbulent premixed flame fronts have
been investigated and considered  for modelling
sub-grid scales in the Large-Eddy-Simulation framework.
Since the width of such thin reaction fronts cannot be resolved into the 
coarse mesh of LES, the extent of wrinkled
flame surface contained in a volume is taken into account.
The amount of unresolved flame front is estimated via the ``wrinkling factor''
that depends on the definition of a suitable
fractal dimension and the scale at which the fractal scaling is lost, the inner
cut-off length $\epsilon_i$.
In this context, the present study considers laboratory experiments and one-step 
reaction DNS of turbulent premixed jet flames in different regimes
of turbulent premixed flames. 
Fractal dimension is found to be substantially constant and well below that 
typical of passive scalar fronts. The inner cut-off length shows a clear scaling
with the dissipative scale of Kolmogorov for the regimes here considered.
These features have been exploited performing Large Eddy Simulations. Good model performance   
has been found comparing the LES against a corresponding DNS at moderate Reynolds number 
and  experimental data at higher Reynolds numbers.

\end{abstract}

\begin{keyword}


LES, Turbulent Premixed Combustion,  Fractal scaling, OH-LIF  
\end{keyword}

\end{frontmatter}


\section{Introduction}

Turbulent premixed combustion at high Reynolds numbers
is characterized by the interplay of
hydrodynamic flow field, heat release and pressure waves,
see e.g. \cite{driscoll2008turbulent,lipatnikov2012fundamentals}.
This intrinsic coupling leads to a substantial large scale unsteadiness that
can be captured by the time resolving feature of Large Eddy Simulation (LES), 
making this approach extremely appealing. 
LES computes explicitly the time dependent dynamics of the large-scale 
structures of the flow, modelling the effects of the unresolved small scales, 
the so-called sub-grid scales. In premixed combustion, the success of the approach 
crucially depends on the ability of modelling the interaction between chemistry and 
turbulence, which occurs in thin reaction regions below the resolved scales.

Several different approaches are described in literature.
A possible methodology consists in artificially thickening the reaction zone, 
to make it resolved on the computational mesh.
Such artificial thickening modifies the interaction between turbulence 
and chemistry and requires the introduction of a suitable efficiency function 
involving several empirical parameters to recover the correct behaviour,
\cite{Meneveau,colin2000thickened}.
Alternatively, the propagative nature of the flame front can be exploited in 
the context of so-called G-equation methods, where
the flame front is represented by a particular iso-surface of a continuous 
scalar field $G$ moving with the given speed of the front,
\cite{williams1985turbulent,peters1999turbulent,pitsch2005consistent,
knudsen2008dynamic,moureau2009level}.

Sticking to classical sub-grid formulations for turbulent stresses and 
transport, the specific combustion closure we pursue in this work is inspired 
to the multiscale nature of turbulent flame fronts, \cite{nicolleau1994eddy,kniveymen}. 
When the area of an instantaneous flame front surface is measured at 
increasingly finer resolutions, the measure increases, following a power law 
with exponent $D$. The exponent is typically non-integer, smaller than the
dimension of the embedding space ($D_E = 3$) and larger than the topological 
dimension of a smooth surface ($D_T = 2$), thus characterising the front as a 
fractal, \cite{mandelbrot1982fractal}. 
In fact, the area increase follows the fractal power law until an 
{\it inner cut-off} is eventually reached. For finer resolutions the scaling 
is lost and the smooth behaviour is recovered. In the technical literature this 
kind of objects are called pre-fractals, \cite{sreenivasan1986fractal}.

In the context of fractal approaches for premixed combustion modelling, a
simple choice consists in describing the reactions 
in terms of a single progress variable 
$c$, determined either from temperature $T$ or fuel mass-fraction $Y_R$, 
e.g.  $c = (T-T_u)/(T_b-T_u)$, with the subscripts $b$ and $u$ denoting 
burned and unburned states, respectively.
The progress variable evolves according to an advection/diffusion/reaction 
equation, where the reaction term $\dot \omega$ is often given by a global 
Arrenhius law.  Similarly to G-equation methods, the propagative nature of the 
front can be exploited to recast the diffusion and reaction terms into a single
propagative contribution characterised by a suitable phase speed, $S_L$,
corresponding to the unstrained Laminar flame speed, \cite{veynante2002turbulent},
\begin{align}
\label{eq:0} 
{\bf \nabla} \cdot   {(\rho D \nabla c)}+{\dot{\omega}}=
\rho_{u} S_{L} \vert\nabla {c}\vert~ \ .
\end{align}
Here $\rho$ is the density, $\dot{\omega}$ the reaction rate and $D$ the 
molecular diffusivity.
In this framework, LES involves the convolution of the RHS of eq.~(\ref{eq:0})
with a filter of width $\Delta$. The result is expressed in terms of flame 
surface density $\Sigma=\overline{\vert\nabla c}\vert$, with the overline 
denoting filtering. $\Sigma$ represents the  convoluted front area contained 
in a volume of dimension $\Delta$.  A number of closures for $\Sigma$ has been 
proposed in literature, starting with simple algebraic expressions originally 
introduced by \cite{boger1998direct}. More directly related to our present 
interest is the link with the fractal properties of the front, 
\cite{kniveymen}. 
In this case, the model exploits the self-similar fractal nature of progress 
variable isosurfaces that are wrinkled and folded by turbulence over a wide 
range of scales, from the integral length scale down to an inner cut-off 
$\epsilon_i < \Delta$.
The wrinkling factor 
$\Xi=\overline{\vert\nabla c}\vert/\vert\nabla \overline{c}\vert$
relates the flame surface density to the gradient of the filtered progress 
variable $\overline c$ and 
corresponds to the ratio of the two areas measured at scales 
$\Delta$ ($A(\Delta)$) and $\epsilon_i$ ($A(\epsilon_i)$), respectively.

As already mentioned, the measure of the front surface scales with a power 
law whose exponent, namely the fractal dimension $D$, is known to be constant 
down to $\epsilon_i$.
More in details, the filtered progress variable equation reads
\begin{align}
\frac{\partial \overline{\rho}\tilde c}{\partial t} & + {\bf \nabla} \cdot (\overline{\rho}\tilde {\bf{u}}\tilde c)+
{\bf \nabla} \cdot {(\overline{\rho}\widetilde {{ \bf{u}} c}-\overline{\rho}\tilde {\bf{u}}\tilde c)}= 
{\bf \nabla} \cdot   \overline{(\rho D \nabla c)}+\overline{\dot{\omega}}=
\label{eq:1} 
\rho_{u} S_{L} \Xi \vert\nabla \tilde{c}\vert~,
\end{align}
where $\overline{\cdot}$ and 
$\tilde{\cdot}$ denote Reynolds and Favre averaging, respectively, and
$\bf{u}$ is the velocity.
Following \cite{kniveymen,kniveymenPOF}, the model can be closed using the fractal behaviour 
of flames 
\begin{equation}
\Xi({\Delta})\propto (\Delta/ \epsilon_i)^{D-2} \ .
\label{eq:1b}
\end{equation}
After properly identifying the inner cut-off $\epsilon_i$, this approach 
circumvents the problem found in LES {where} reactive processes act in the 
unresolved range of scales, below the LES filter width.

Coming back to our purposes, a crucial issue of this fractal model is the determination of the
fractal dimension $D$ and of the inner cut-off $\epsilon_i$ and their dependence on
turbulence/chemistry conditions.

Some studies investigated the fractal features of premixed flames, i.e.\ the fractal dimension, the 
inner and the outer cut-offs, see for example~\cite{cohhalchagokgul,chahawaspkerkolche}.
The main aim is to determine the dependence of these parameters on turbulence, thermodynamics and 
flame features.\\
Pioneering investigations on the fractal features of interfaces passively transported in non reactive turbulent 
flows as wakes beyond bodies, jet flows or mixing layers can be found in~\cite{sremen1986,srerammen}.
The experimental findings indicating a fractal dimension of 2.37 and a cut-off length proportional to the 
Kolmogorov dissipative length was also supported by an interesting theoretical reasoning based on the 
Kolmogorov K41 theory and intermittency corrections. However the {mechanism} producing a surface 
wrinkling in cold (incompressible) turbulent flows is deeply different from what happens to the flame front 
wrinkled by the turbulent structures where the thermal expansion due to the combustion induces 
additional phenomenologies.
Actually, concerning turbulent premixed flames, the fractal dimension appears to show a wide variation between 2.18-2.35, 
see e.g.~\cite{gul,gulsmawonsnesmidessau}, and the inner cut-off length dependence is still debated~\cite{gulsma}. 
The outer cut-off scale is generally accepted to be grater than the turbulence integral scale~\cite{gulsmawonsnesmidessau},
however it should be noted that this information is not needed in the LES framework.
\cite{cheman2003} reports a geometric interpretation on fractal flame features analysing  
data of a turbulent premixed Bunsen flame. They illustrate the relations between the inner and outer cut-off 
length with the turbulence and flame properties.
The study in~\cite{gulsmawonsnesmidessau} presents an extensive experimental analysis of a premixed propane/air 
flame using OH-LIF and Mie-Scattering, with different turbulence intensities $u'/S_L$ and Reynolds numbers. The paper
discusses also the possible errors induced by experimental image acquisition technique showing a strong dependency
on the employed methodology and providing technical explanation of observed differences.
Furthermore the authors remark as the large discrepancy between predictions and previous studies
calls for new investigations.\\
To partially 
overcome these difficulties a dynamic fractal model for LES has been developed by~\cite{kniveymen,kniveymen2004} exploiting
the self-similar nature of the process. The model was tested using a-priori analysis on 
 experimental data concerning turbulent premixed propane/air flame stabilised with a triangular bluff 
body. Data show a good agreement between the a-priori model prediction and the experimental data.

Aim of the present work is twofold, on one hand the experimental and numerical evaluation of  both the exponent $D$ and $\epsilon_i$ by
OH-LIF --Laser Induced Fluorescence-- on premixed
methane/air jet flames in round and annular configurations.
On the other hand, on the ground of the fractal characteristics found, we aim to accurately compare the Large-Eddy-Simulation 
data with a corresponding DNS of a
turbulent premixed flame  at moderately low Reynolds number, i.e.\  
$Re=U_b D \rho/\mu = 6000$ ($U_b$ is the bulk 
jet velocity and $D$ is the nozzle diameter) and to reproduce
the main features of the higher Reynolds number  experiments ($Re=16000$ and $Re=24000$) carried out by \cite{cheman}.

Results show that fractal exponent $D=2.23$, smaller than in non-reactive cases, 
appears to be rather insensitive to the explored Reynolds numbers, 
turbulence and flame features  and to the configurations adopted. 
It follows that fractal dimension can be considered as a fixed parameter for turbulent 
combustion processes. The experimental analysis on the inner cut-off $\epsilon_i$ shows that  
it scales with the dissipative Kolmogorov length $\eta_k$ in the regimes here analysed. 
The results of the Large-Eddy-Simulations using these assumptions  well agree with the corresponding reference data. 

\section{Fractal characteristics of flame fronts}

The fractal features of the flame front of turbulent premixed flames are extracted analysing
experimental and numerical (DNS) data. The dataset pertains to a wide range of flame regimes
as can be appreciated in  figure~\ref{fig:borghi} where a turbulent flame classification diagram is reported
with the present experiments (closed symbols) and DNS (open symbols).

\subsection{Methodology}
\subsubsection{Experimental set-up}

\begin{figure}
\centering
\includegraphics[width=.75\textwidth]{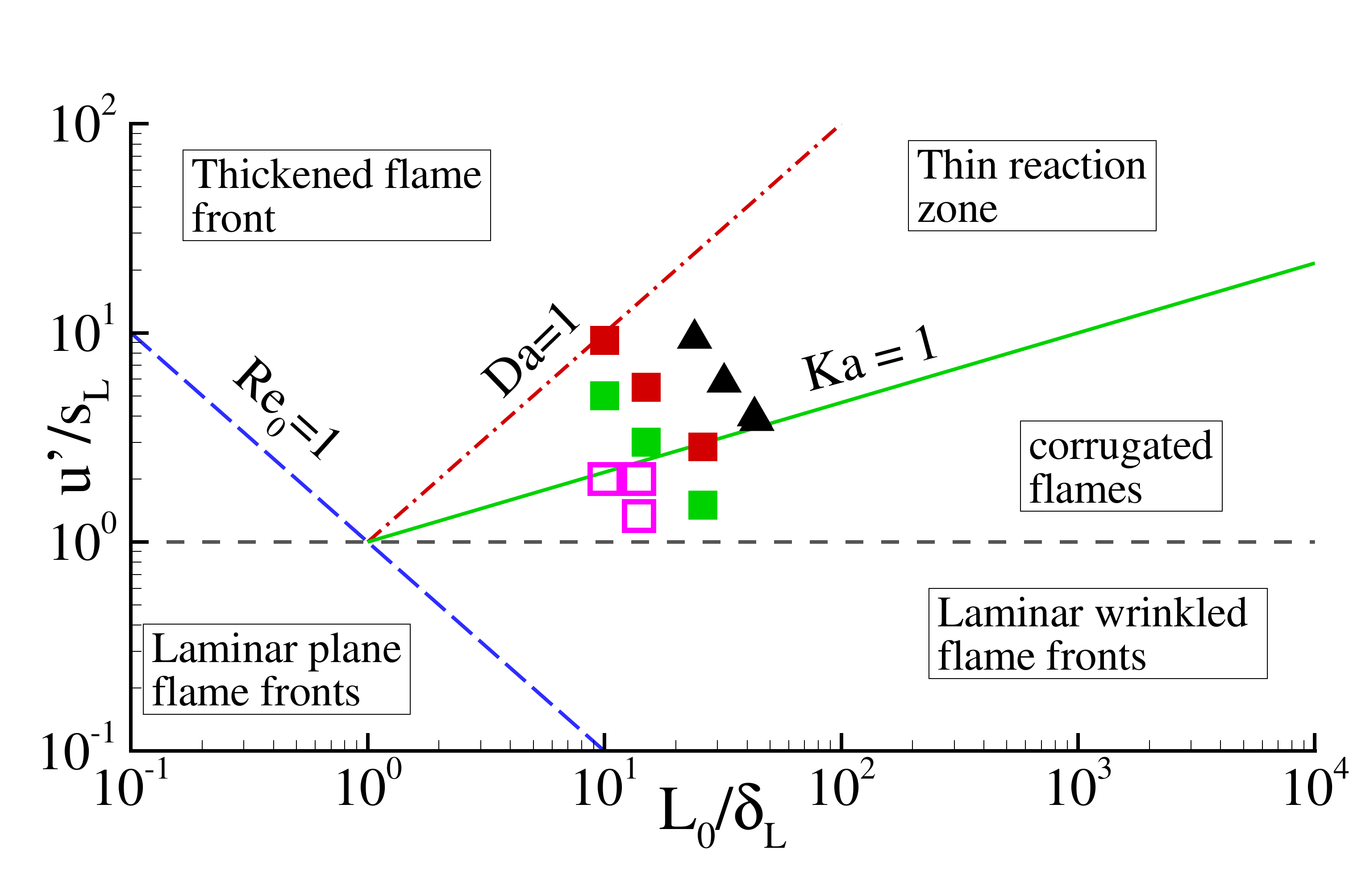}
\caption{ Right panel: Turbulent premixed flame classification diagram (Borghi)  
where present Experiments (closed symbols: squares round Bunsen jet at $Re=7000$ and $10000$, triangles
annular jet at $Re=10000$) and DNS (open triangles $Re=4000$ and 6000) flame regimes are highlighted.
In the figure, $u'$ is the typical level of velocity fluctuations, $L_0$ the integral flow scale, 
$S_L$ and $\delta_L$ the laminar flame speed and thickness, respectively. $Ka=(\delta_L/\eta_k)^2\sim
(u'/S_L)^{3/2}\, (L_0/\delta_L)^{-1/2}$ is the Karlovitz number ($\eta_k$ the
Kolmogorov length) and $Da=S_L \, L_0/(u'\, \delta_L)$ is the Damk\"oler number. 
\label{fig:borghi}}
\end{figure}

Methane/air premixed flames have been realised in two different jet-burners.
The first is a cylindrical bunsen burner, whose inner diameter is $D = 18$ mm.
A diffusive methane pilot flame anchored the flame to the nozzle exit.
Reynolds numbers (diameter based, i.e. $Re=U_b D/\nu$) and equivalence ratios ($\Phi$) ranges between $Re=7000$ and 
$Re=10000$ and from stoichiometric mixtures down to $\Phi = 0.6$ \cite{tro2009}.
The second is a stainless steel annular inlet bluff-body stabilised burner 
{(nozzle outer diameter $D=25$ mm, inner 
diameter $D_i=15$ mm)}
fed with a mixture of $CH_4$ and air at different equivalence ratios.
The Reynolds number evaluated by the bulk velocity and the nozzle diameter
 is kept fixed at a value of $Re = U_b D/\nu=10^4$, to ensure a fully 
developed turbulent jet with well-defined turbulence characteristics 
\cite{tromargiaca}.
In this case the flame is anchored by the recirculation due to a conical
bluff-body. Four different flames are investigated and their methane/air
equivalence ratio $\Phi$ is varied in the range of $[0.6-1.12]$.
Velocity data comes from LDA and PIV measurements by seeding the 
flow with $5~\mu$m alumina particles \cite{picbattrocas,troiani2013turbulent}. The flame classification
of the present experiments spans from the beginning of the thickened flame to the {corrugated} flame regimes, as reported in figure~\ref{fig:borghi}.

Flame front detection is performed instead by the acquisition of fluorescence 
signal emitted by OH radicals.
To that end, a Nd:YAG laser beam is delivered through a tunable dye laser 
coupled with a second-harmonic generator crystal in order to shift the 
laser wavelength from $532$ nm down to $282.93$ nm, corresponding to the 
$Q_1(6)$ absorption line of OH.
A suitable cylindrical lens expands the beam into a $350~\mu\rm{m}$ thick 
laser sheet.
The resulting OH fluorescence emission is around $309$ nm and is then 
collected by a $1024 \times 1024$ pixels ICCD ($2 \times 2$ pixel binning) 
equipped with a $78$ mm Nikon quartz lens, resulting in a map of 
$512 \times 512$ equivalent pixels with a resolution of $160~\mu\rm{m}$ 
for each equivalent pixel.
Furthermore, a narrow pass-band filter, $10$ nm wide and centered at 
$310$ nm, isolates the relevant spectral line.
\begin{figure}[h!]
\centering
\includegraphics[height=.40\textwidth]{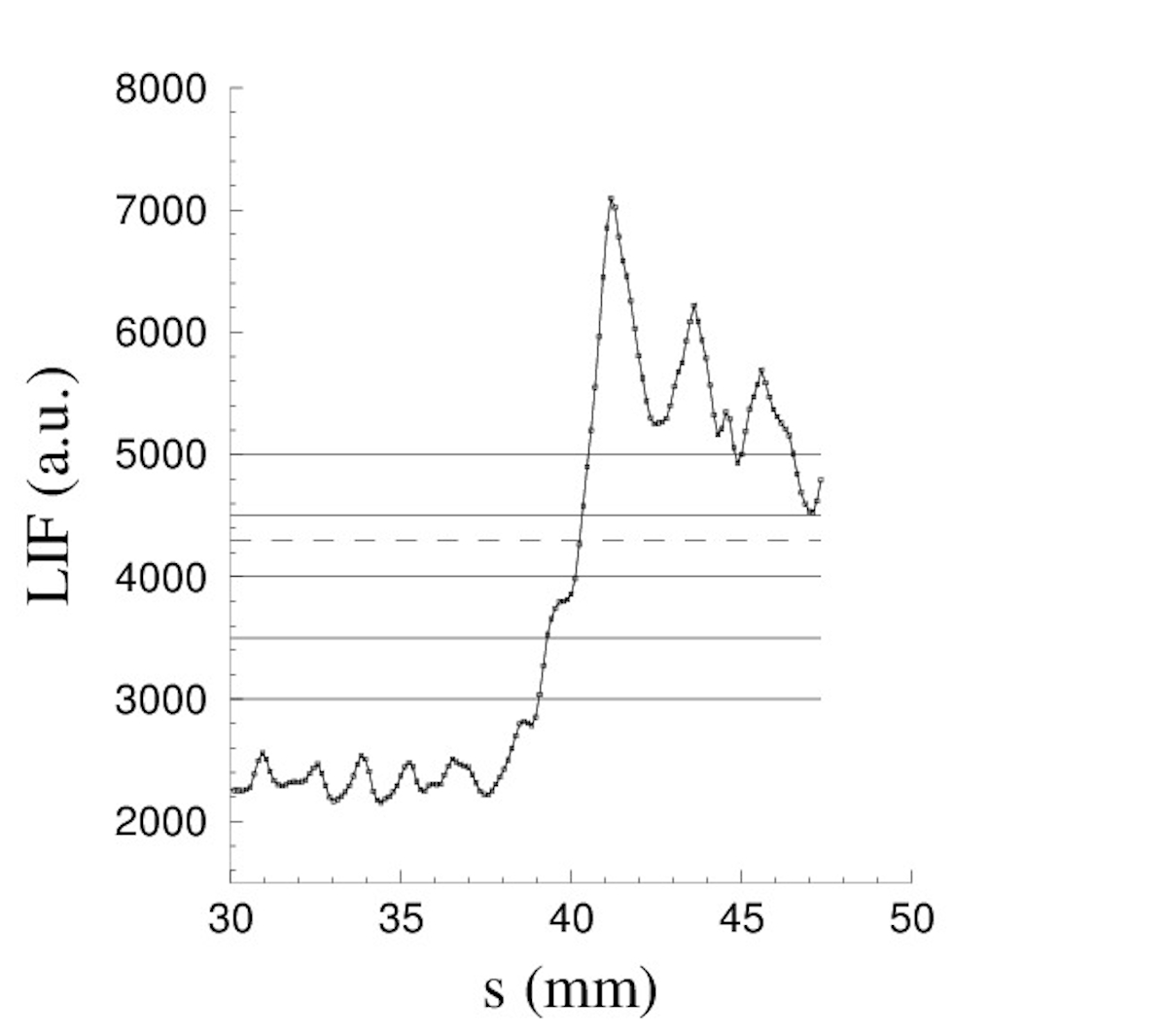}
\hfill
\includegraphics[height=.40\textwidth]{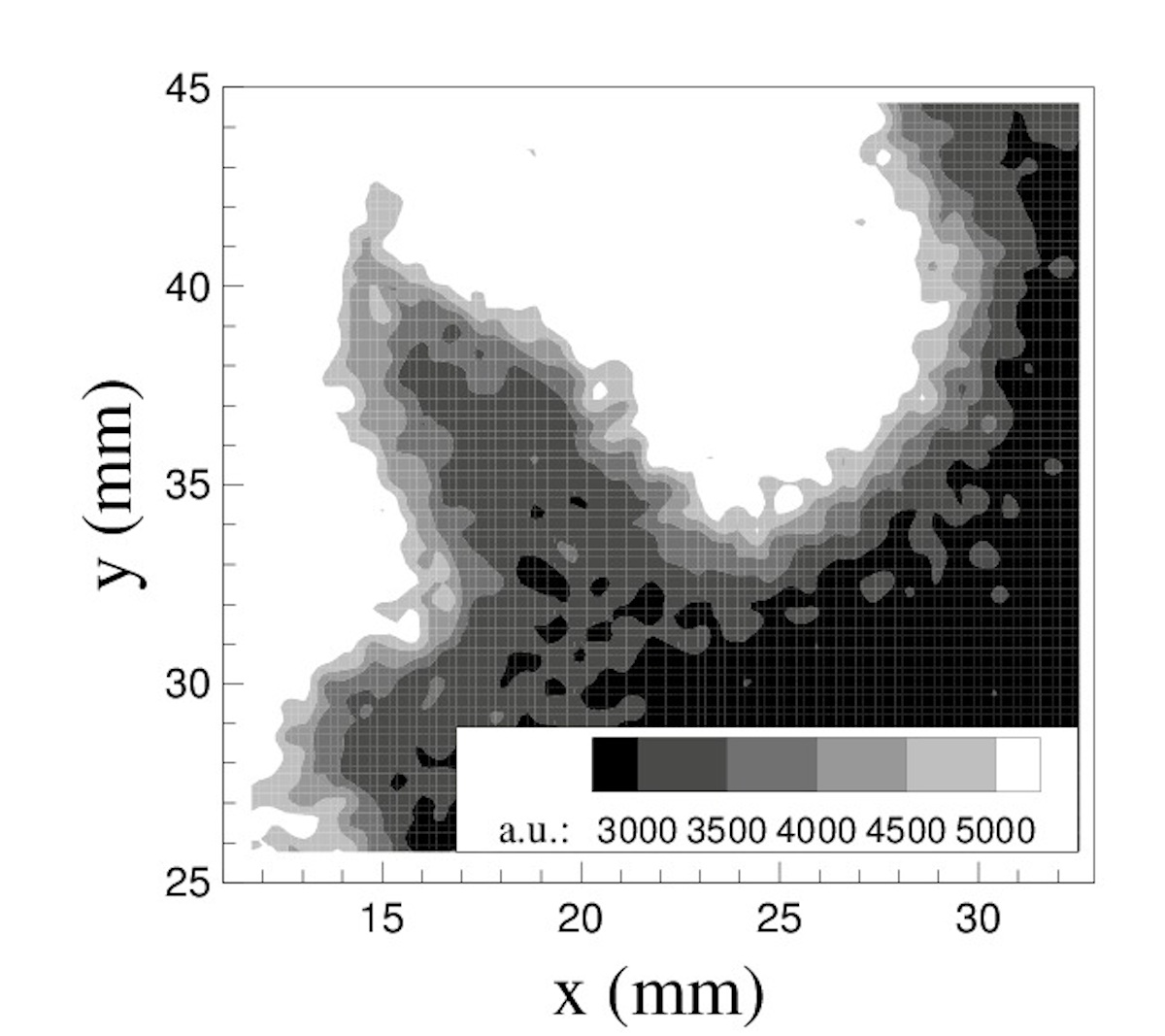}
\caption{Left: OH fluorescence signal, $\Phi = 1.03$.
Dashed line corresponds to maximum gradient position.
Abscissa $s$ originates at a point arbitrarily chosen in the reactants and
crosses the flame front normally towards combustion products.
Right: OH fluorescence map, $\Phi = 1.03$. Gray levels refer to different OH
threshold values, as reported in the legend.}
\label{fig:OH_cut}
\end{figure}

The fluorescence signal from OH radical is proportional to its concentration 
and relative measurements are possible (absolute measurements of 
concentrations are instead prevented from non-radiative disexcitation channels 
that are generally active together with detectable fluorescence emission).
An example of the LIF images taken with the experimental set-up described above 
is given in the right panel of figure~\ref{fig:OH_cut}. 
Here, the image portrays a detail of the whole flame and the signal level is 
reported in terms of CCD counts whose gray-color scale is shown in the 
inset appearing at the bottom.
Moving from the zones in white color representative of higher level of 
fluorescence towards darker regions, it appears that the fluorescence signal 
decreases abruptly across the flame front, where the OH radicals are formed.
This typical behaviour is shown in the left panel of figure \ref{fig:OH_cut} 
where a cut perpendicular to the flame front has been extracted from the image.
Given that, within the products OH radicals disappear at a much slower rate 
than that characterising their formation in the flame, the consequent 
asymmetric behaviour can be used to distinguish between reactant and product 
zones.
Based on this result, it is worth emphasising that even though in our 
measurements the signal-to-noise ratio is usually sufficiently large
the gradient field may suffer from significant high-frequency contamination.
All the same, the signal may be treated by suitable filtering techniques, e.g.,
the non linear filter described in \cite{perona1990scale}, which provides 
satisfactory results in this case.
Nonetheless, a variant of the threshold method \cite{kalt2002experimental} 
is adopted, which locates the front at the isoline that better correlates 
with the maximum gradient in the region of interest.
\begin{figure}
\centering
\includegraphics[height=.45\textwidth]{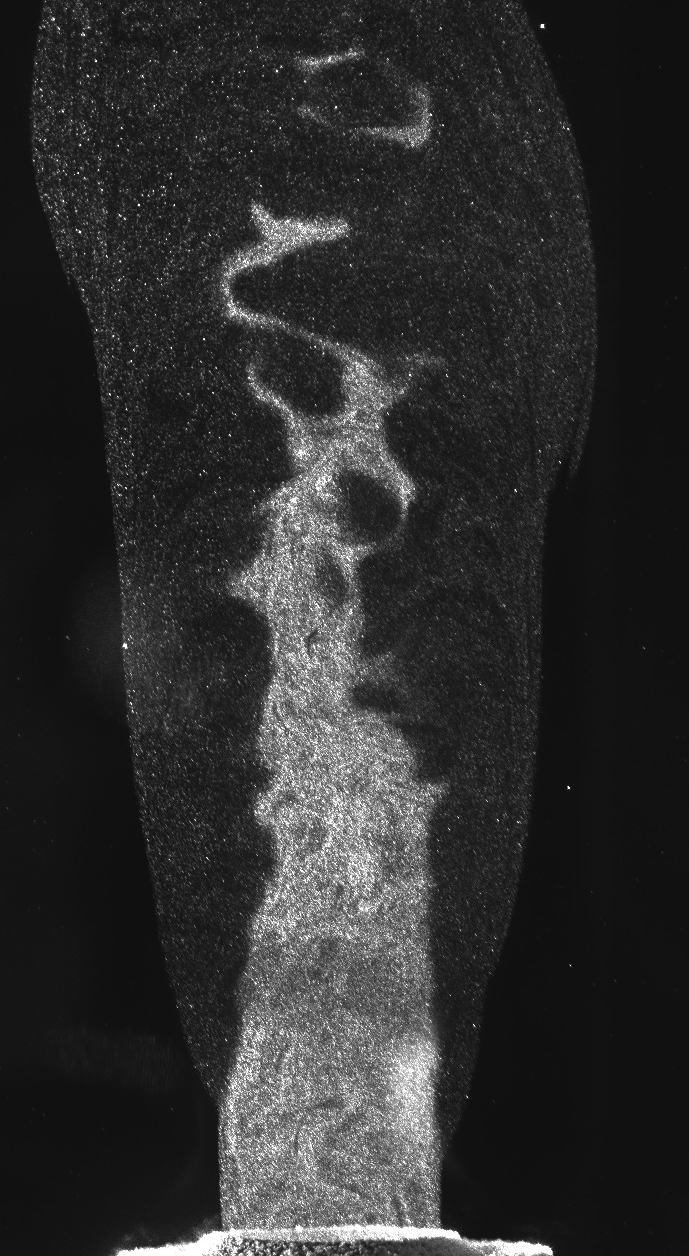}
\hfill
\includegraphics[height=.45\textwidth]{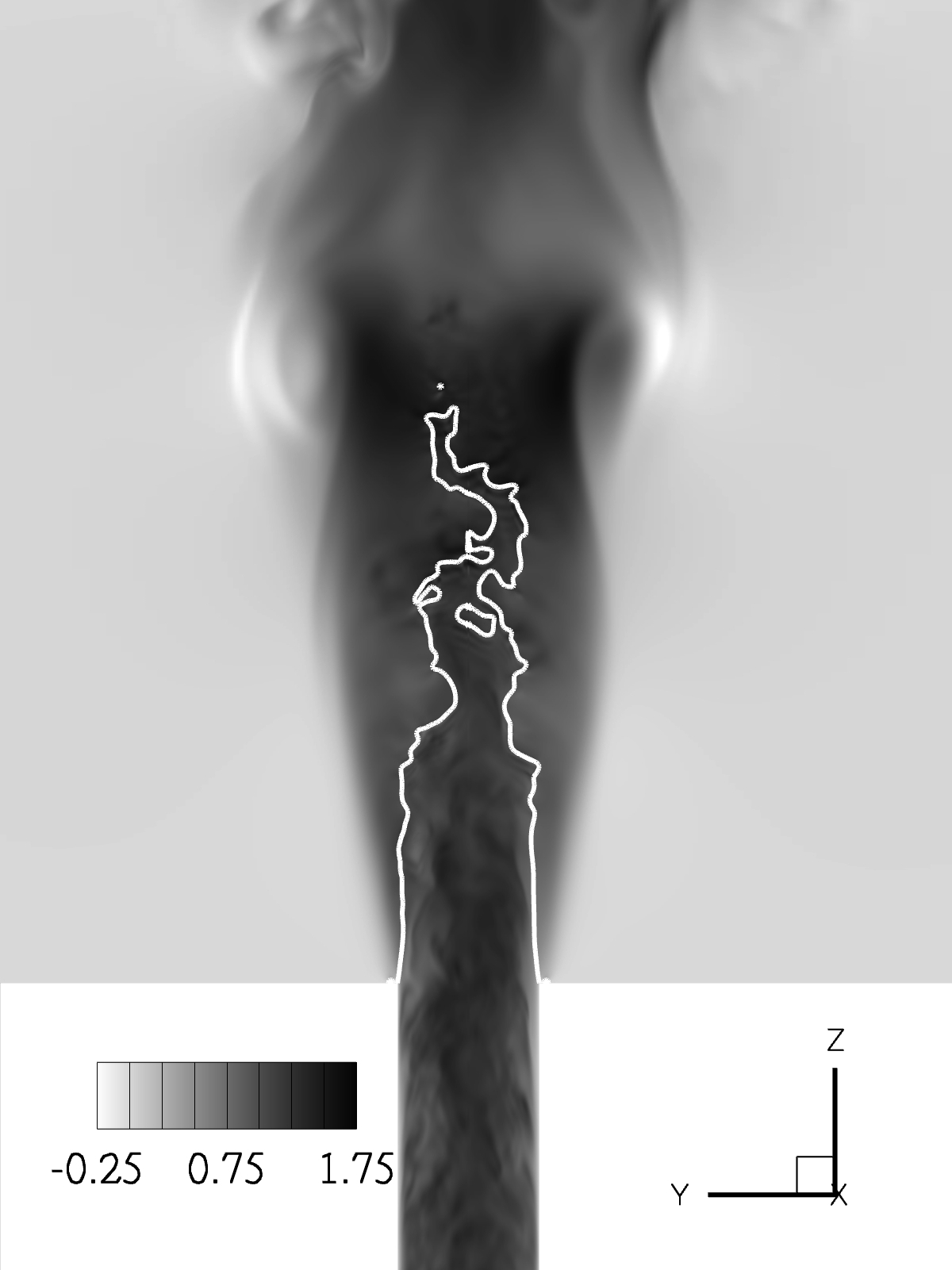}
\hfill
\includegraphics[height=.45\textwidth]{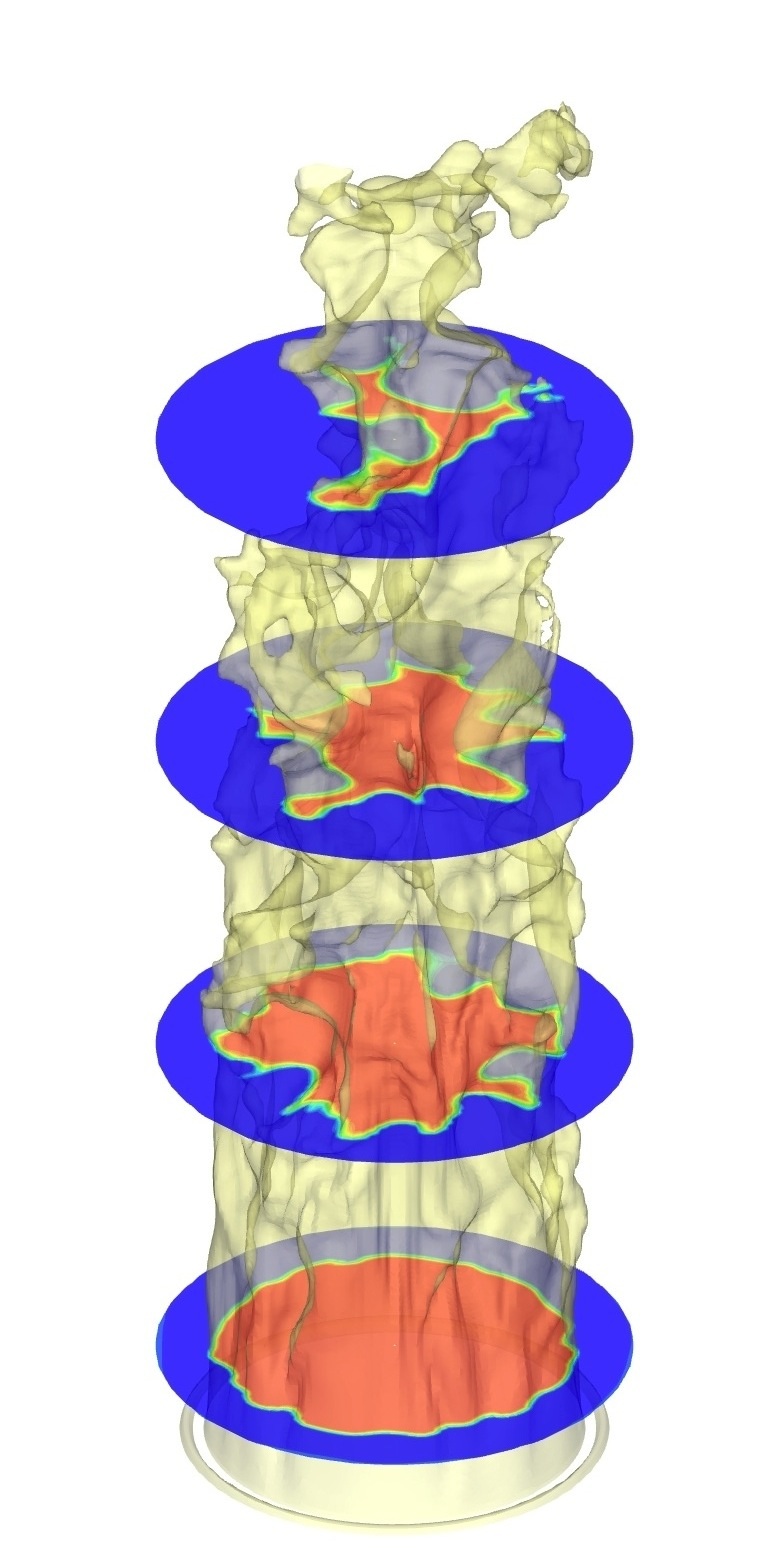}
\caption{Snapshots. Left panel, 
raw mie scattering of the Bunsen flame at $Re=6000$;
middle panel, {DNS of a Bunsen flame at $Re=6000$:} 
contours of axial velocity with superimposed a isolevel of
$c=0.5$ in white; right panel, instantaneous 3D flame front detected by means of
the reactant based progress variable isolevel $c=0.5$, four planes are superimposed to highlight
the flame wrinkling.
\label{fig:1b}}
\end{figure}
%
%
\subsubsection{Numerical Algorithm}

The direct numerical simulation algorithm discretises the cylindrical formulation of the
Low-Mach number asymptotic {non-dimensional} Navier-Stokes equations~\cite{majda}, 
%
\begin{align}
\label{eq:cont_def}
&\frac{\partial \rho}{ \partial t} + \nabla \cdot (\rho {\bf u}) = 0  
\\
\label{eq:mom_def}
&\frac{ \partial \rho \bf u}{\partial  t} + \nabla \cdot ({\rho \bf u \otimes \bf u}) =
\frac{1}{Re}\nabla \cdot {\bf \Sigma} - \nabla P +\rho {\bf g} \\
\label{eq:spec_def}
&\frac{\partial \rho c}{\partial  t} + \nabla \cdot ({\rho {\bf u }c}) =
\frac{1}{Re Sc} \nabla \cdot (\rho {\cal D} \nabla c) + \, \dot{\omega} \\
\label{eq:div_def}
&\nabla \cdot {\bf u} =\frac{1}{p}\left[ 
\frac{1}{Re Pr } \nabla \cdot (k \nabla T)
 + \frac{\gamma -1}{\gamma}  Ce \,\dot{\omega} \right]\\
\label{eq:state_def}
&T=\frac{p}{\rho}
\end{align}
%
where ${\bf \Sigma}=2\,\mu \,{\bf S}=\mu\,(\nabla \bf u + {\nabla \bf u}^{T})$ is the 
viscous stress tensor and $\mu$ is the viscosity dependence on temperature, 
$\rho$, $\bf u$ and $P$ are the density, the velocity and the 
dynamic pressure, respectively;  $T$, $p$, $c=1 - Y_R/Y_{R0}$ and $\dot{\omega}$  
are the temperature, the thermodynamic pressure (constant in both space and time), 
the progress variable defined in terms of the local $Y_R$ and the inlet $Y_{R0}$ reactants concentration,
and the corresponding reaction rate, respectively. 
${\cal D}$ and $k$ are the progress variable diffusivity and the thermal conductivity, respectively.
$Re=U_b D \rho_0/\mu_0$ is the Reynolds number, with $U_b$ 
the inlet bulk velocity, $D$ the nozzle diameter, and the subscript $0$ indicating the reference quantities 
corresponding to jet inlet conditions,  $Pr=\mu_0/(k_0 c_{p0})$ and $Sc=\mu_0/(\rho_0 D_0)$ 
are the Prandtl and the Schmidt numbers giving the ratio between thermal and mass diffusions 
on the viscosity $\mu_0$, respectively, here $c_{p0}$, $k_0$, and $D_0$ are the heat capacity at constant pressure, 
the thermal diffusivity and the mass diffusion evaluated at the reference state; $\gamma=c_p/c_v$ and 
{$Ce=\rho_0 \Delta H/p_0$ are the heat capacity coefficient ratio and the heat release parameter,
respectively with $\Delta H$ the enthalpy of the reaction.}

Spatial discretisation is obtained by central second order finite differences
in conservative form on a staggered grid. The convective terms of scalar equations are discretised
by a Bounded Central Difference Scheme to avoid spurious oscillations, see~\cite{watdec} for details. 
Temporal evolution is performed by a low-storage third order Runge-Kutta scheme.  
\\%
Time-evolving boundary conditions are prescribed for the
inflow where a fully turbulent inflow velocity 
is assigned at each time-step by using a cross-sectional slice of a fully
developed pipe flow obtained by a companion DNS. 
In the middle panel of figure~\ref{fig:1b} the discharge of the fully turbulent pipe flow
creating the jet
is shown by contours of the fluctuating axial velocity.
The density and the
concentrations are kept fixed on the inflow nozzle. The outer part of the inlet region
is constituted by an adiabatic wall.  
A convective Orlanski condition is adopted for outflow of
all the variables, see also~\cite{orla,boersma,pic_cas}. The artificial side boundary is modelled by an adiabatic
 traction-free condition to allow a correct entrainment rate.
More details on the code and tests for incompressible jets can be found in 
\cite{picbattrocas,pic_cas,picsarcas,batpictrocas}.
\\%
\begin{figure}
\centering
\includegraphics[width=.46\textwidth]{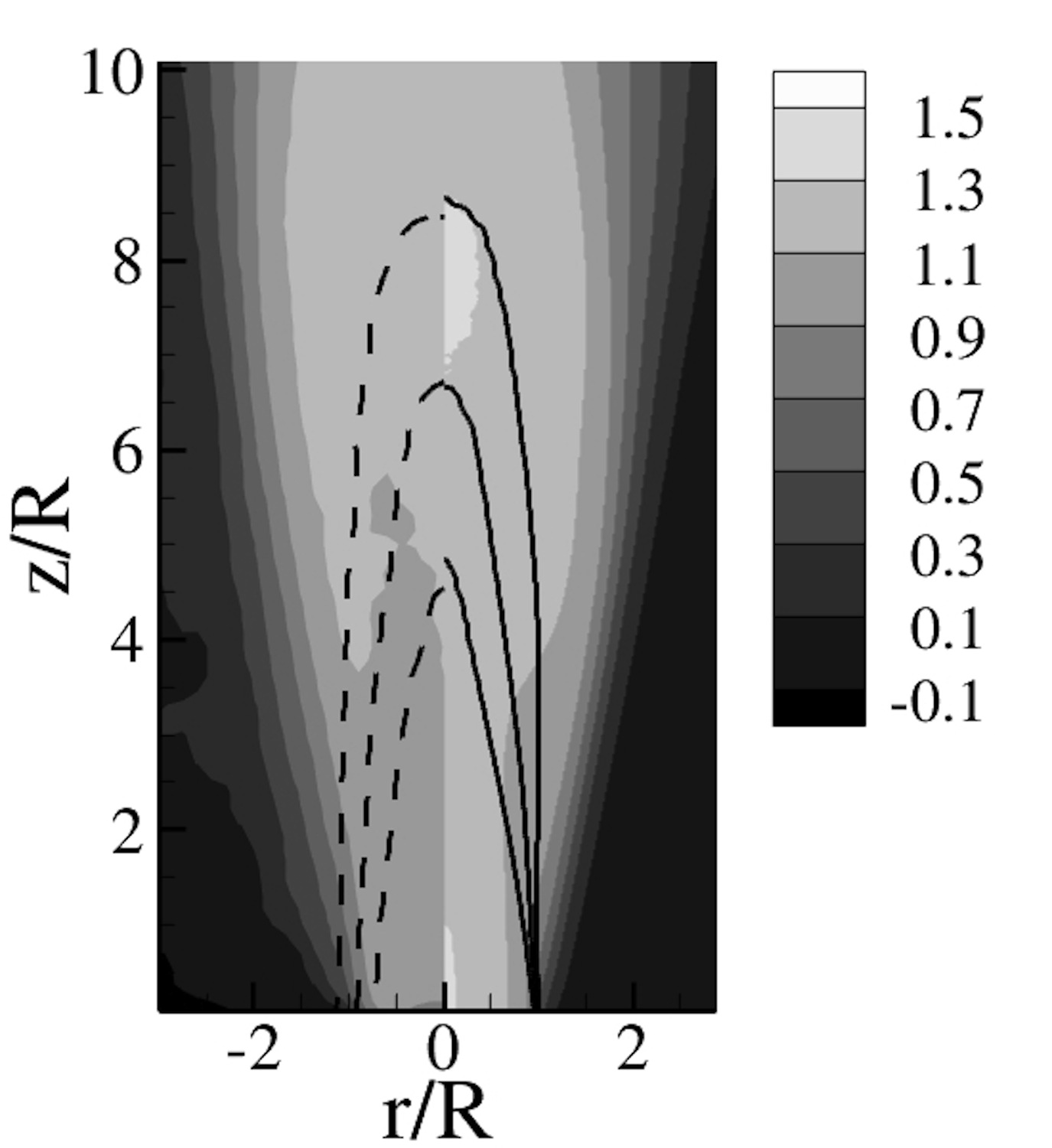}
\hspace{.5cm}
\includegraphics[width=.46\textwidth]{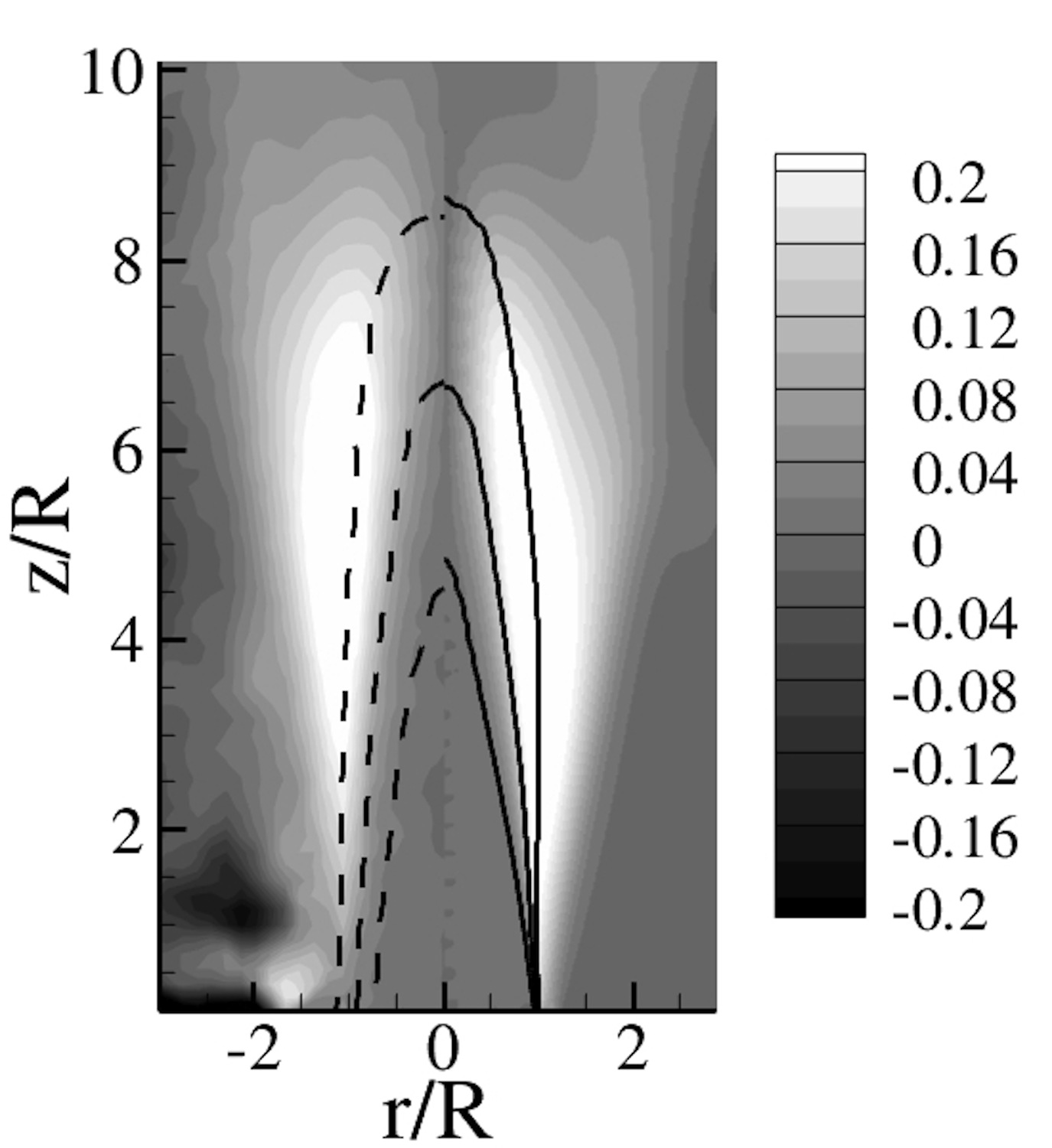}
\caption{Mean flow field for experiments and DNS at $Re\simeq6000$.
Half-figures on the left represents experimental data, right half-figures 
the DNS ones. Right panel, axial mean velocity; left panel, radial mean velocity.
The isolines provides three isolevels of the progress variable $c$: 0.25, 0.5, 0.75, namely.
\label{fig:2}}
\end{figure}
Concerning the direct numerical simulation,
the chemical kinetics is given by an Arrhenius single-step irreversible reaction
which transforms the premixed fresh mixture $R$ into the exhaust gas of
combustion products $P$:
$\dot \omega=\omega^0 (\rho Y_R) e^{-T_a/T}$,
with $\omega^0$ the reaction rate constant and  $T_a$ the activation temperature.

Three DNS of premixed Bunsen jets are performed in order to reproduce
the main features of  premixed methane/air flames in conditions similar to the present experiments.  
The simulations replicate turbulent flames with $T_f/T_0\simeq 5.3$, and 
$S_L/U_b\simeq 0.05\div0.075$ ($u'\simeq0.1\,U_b$) which can be classified according to the diagram
 shown in figure~\ref{fig:borghi}. As can be appreciated the DNSs are close to the straight line given by
 the Karlovitz number $Ka=1$ as usual in DNS simulations because of the constrained spatial resolution. 

Concerning the input parameters assigned to the numerical code to simulate these flames, we have
fixed $\gamma=1.3$ and $Ce=18.5$ in order to have a 
adiabatic flame temperature of {$T_f=5.3\, T_0$}
~\cite{poivey} ($T_0=300\,K$). The activation temperature 
 is assumed {$T_a \simeq 6\, T_f$},  while the the reaction rate constant 
 $\omega^0$ (made dimensionless by $U_b/R$) varies in the range $30000\div
 67500$ together with the Reynolds number $Re=4000\div6000$ in order to change
 $u'/S_L$, i.e.\ $U_b/S_L$ and $L_0/\delta_L$ as reported in figure~\ref{fig:borghi}.  

The flame has unitary Lewis number with $Pr=Sc=0.7$  
while the diffusion coefficients depend on the temperature following the Sutherland-like law,
{$\mu/\mu_{0}=\rho {\cal D}/\rho_0 {\cal D}_0= k/k_0= \,(T/T_{0})^{1/2}$}.

The computational domain is given by:
$[\theta_{max}\times R_{max} \times Z_{max}]=[2\pi \times 12.34\,R \times
14\,R]$ and it is discretised by
$N_\theta\times N_r\times N_z=128\times201\times560$ points with
stretched mesh in the radial direction to assure resolved 
shear layers with a radial grid spacing at the nozzle exit  $\Delta R\simeq 1.5 \eta_{k}$,
with $\eta_{k}$ the pipe flow Kolmogorov length near the wall.\\
The general agreement between the instantaneous snapshots of the experimental 
and of the DNS flame in similar conditions ($Re=6000$, $\omega^0=45000$, $u'/S_L\simeq2$) is remarkable, 
see figure~\ref{fig:1b}.
A  crucial issue for a correct {reproduction} of the experimental configuration is the accurate turbulent inflow 
conditions that avoid the use of synthetic turbulence or the mean advection of a frozen turbulent field.
Actually, inlet conditions of the experiment and of the DNS are quite similar
with the difference that a fully developed pipe is assigned to DNS, while
in the experiments, inlet is generated by a pipe not long enough to provide fully
developed statistics.    
Nonetheless, DNS with laminar inflows (not presented here) are 
characterised  by the presence of large-scale motions influencing the whole 
reactive jet dynamics, whose behaviour is not observed experimentally. 
Figure~\ref{fig:2} shows the mean flow behaviour for DNS and experiments in the
same conditions, {obtained by averaging about one-hundred instantaneous fields provided by DNS 
simulations and PIV-OH/LIF measurements}. The general matching in terms of flame height and mean velocity
is apparent.

\subsection{Fractal scaling of the front}
\begin{figure}[t]
\centering
\includegraphics[height= .4 \textwidth]{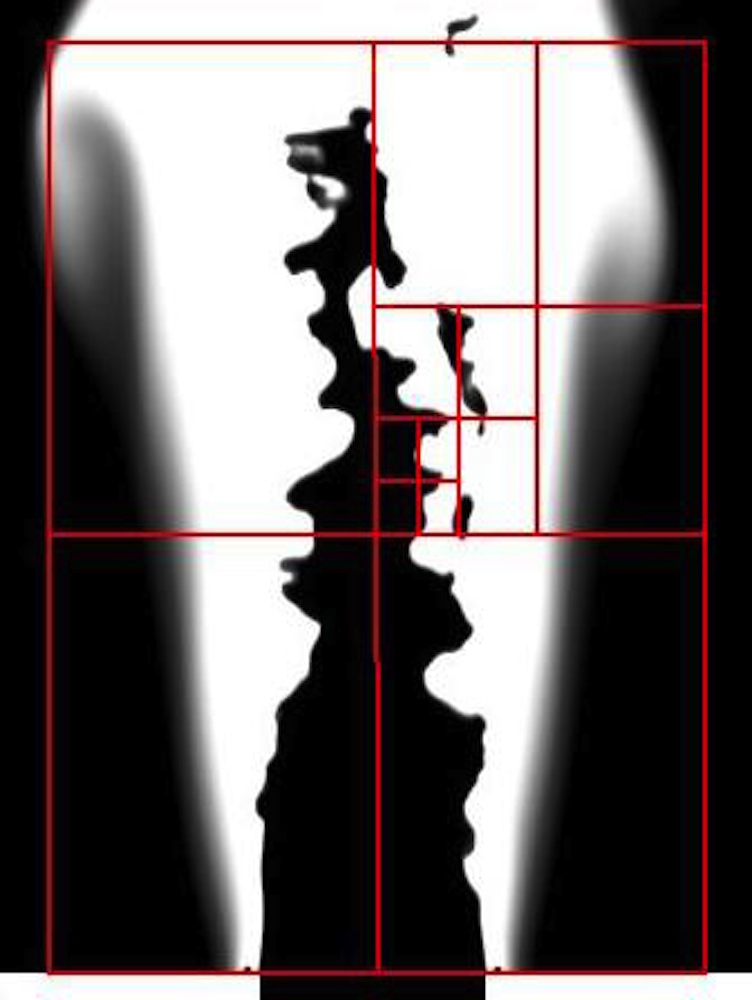}
\hspace{1.cm}
\includegraphics[height= .4 \textwidth]{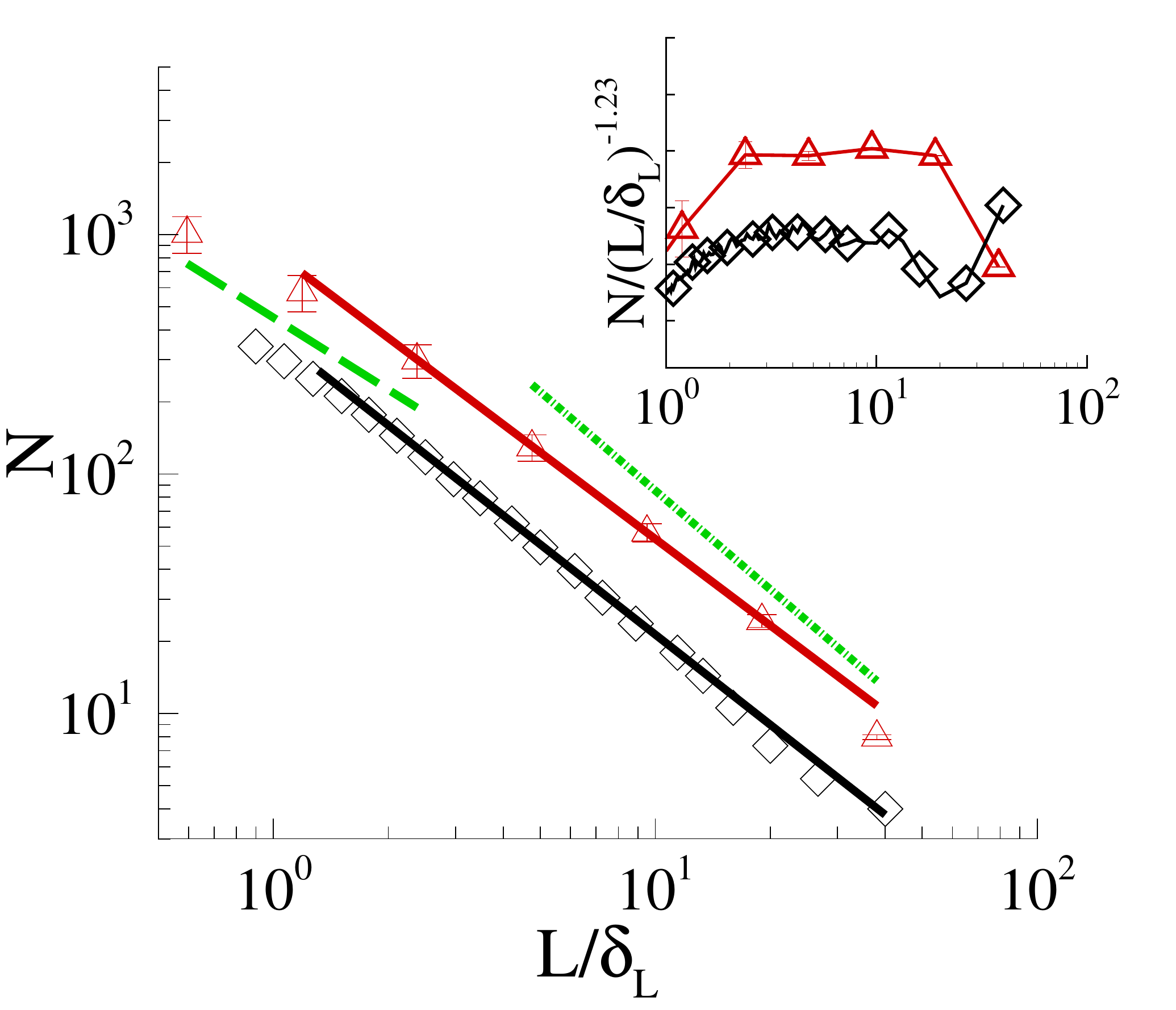}
\caption{Left panel: Example of boxes of decreasing size used to cover the flame front.
The image relates to the DNS data, but the same technique is applied to OH-LIF 
experimental data.
Right panel: ensemble-averaged fractal scaling of 2D-cut flame front by means of
box-counting.
$N$ is the number of boxes at the respective measurement scale $L$.
Abscissa has been made dimensionless by  the laminar
flame front thickness $\delta_L$.
Symbols: Triangles, experiments on cylindrical Bunsen
burner at $Re = 7000$ $\Phi = 1$; diamonds, DNS at $Re = 6000$ and $u'/S_L = 2$. 
Thick lines are linear regressions of each ensemble-averaged data set in
Log-scale ( $N \propto (L/\delta_L)^{-1.23}$).
Dash-dotted line is the theoretical fractal scaling for non-reacting scalar
turbulent iso-surfaces, i.e. $N \propto (L/\delta_L)^{-1.37}$.
Long dashed line represents the viscous closure, $N \propto (L/\delta_L)^{-1}$.
Inset, compensated plot: $N / (L/\delta_L)^{-1.23}$}
\label{fig:4a} 
\end{figure}

\begin{figure}
\includegraphics[width= .48 \textwidth]{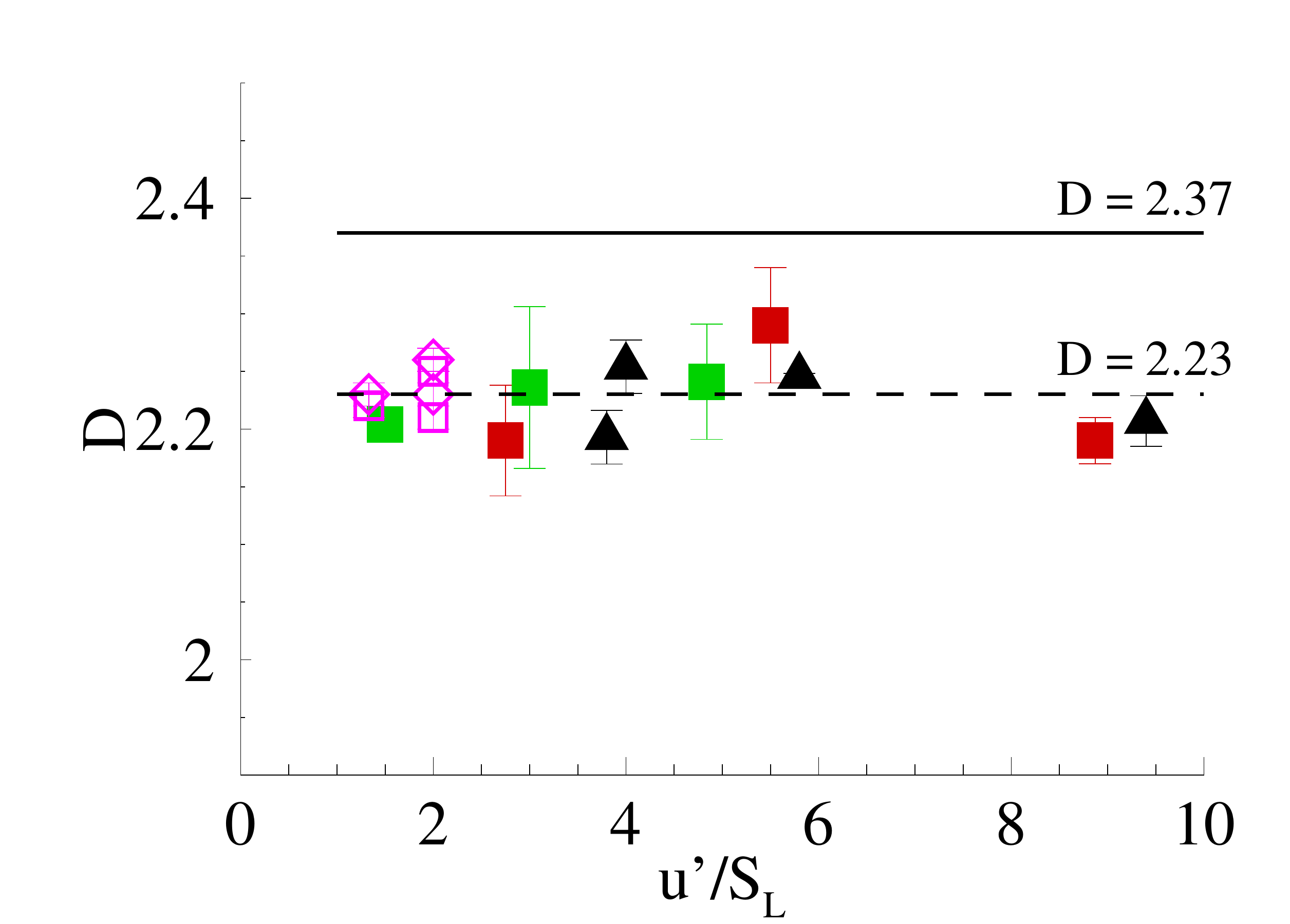}
\hfill
\includegraphics[width= .48 \textwidth]{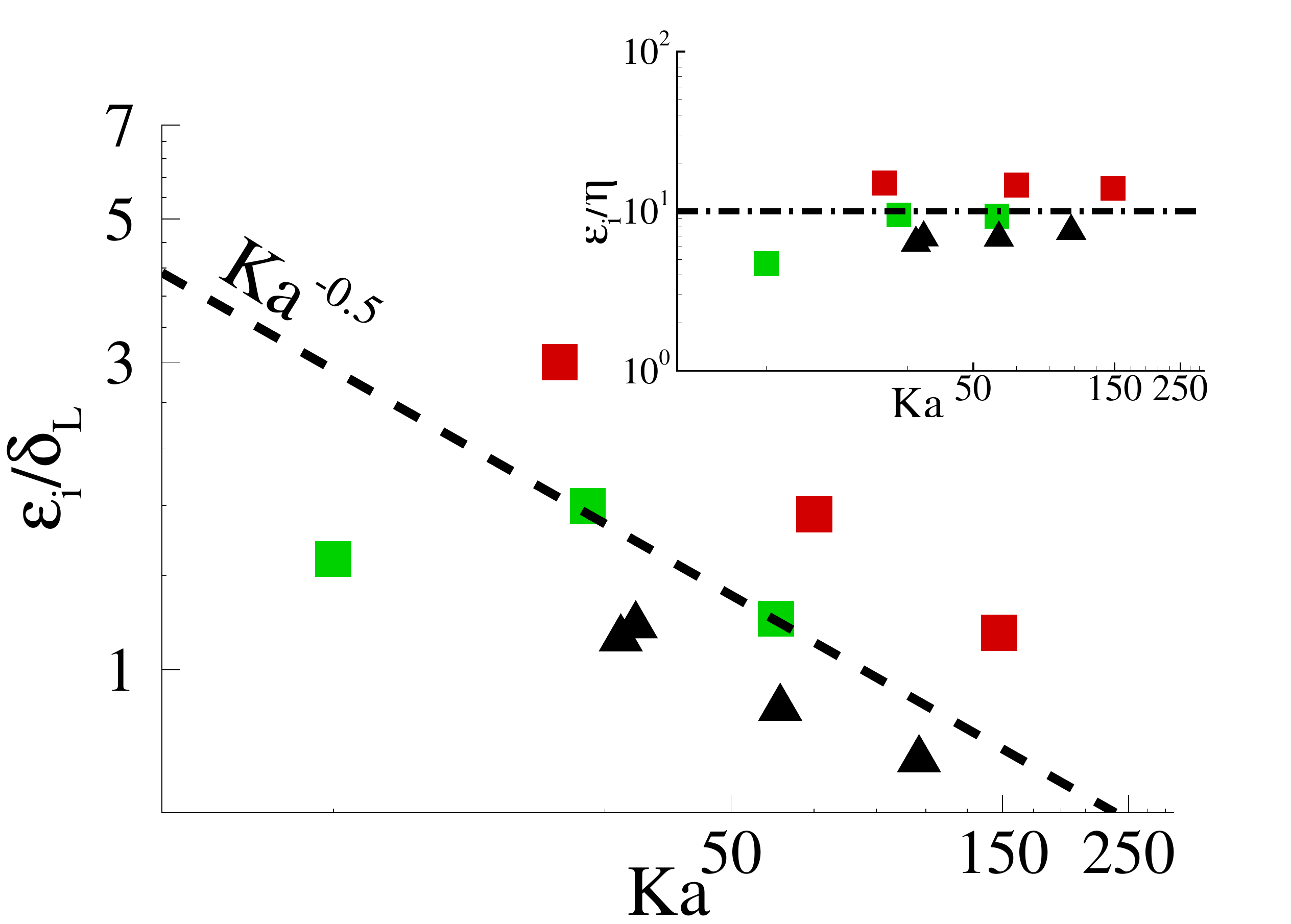}
\caption{ Left panel: fractal dimension D {\it vs} turbulent fluctuation intensity $u'$ normalised
by $S_L$.
Error bars report the fitting error.
Open symbols, DNS: $D$ directly extracted from 3D iso-surfaces ($c=0.5$) by diamond and $D=D_2+1$ extracted from 2D cuts
of the iso-surfaces ($c=0.5$) containing the jet axis by triangles; 
closed symbols, experiments where $D=D_2+1$ is obtained by 2D cuts containing
the jet axis.
Thick continuous line, fractal dimension of non-reacting scalar turbulent
iso-surfaces, i.e. $D = 2.37$. Dashed line, $D=2.23$ present fit of the fractal dimension.
Right panel: inner cut-off ($\epsilon_i/\delta_L$) {\it vs} Karlovitz number $Ka$. Only experimental values
have been reported since DNSs do not cover a sufficiently wide range of $Ka$.
Dotted line represents the scaling $\epsilon_i/\delta_L \propto Ka^{-1/2}$. {In the inset the inner cut-off length 
normalized by the Kolmogorov scale, $\epsilon_i/\eta$, is reported against Karlovitz number. The dash-dotted line 
represent the constant fitting value $\epsilon_i/\eta=10.$}}
\label{fig:1}       
\end{figure}

%
%
The multiscale nature of the surface suggests the application
of geometrical  concepts from fractal theory to quantify the amount of flame wrinkling in terms of
$\Xi$, see equation \eqref{eq:1b}.
In fractal theory, a power law relationship exists between the number of
boxes needed to cover a fractal object and the box size $L$, 
\begin{equation}
N(L) \propto L^{-D}~, 
\label{scal_D}
\end{equation}
where $D$ is defined as the fractal dimension of an object embedded in a three dimensional
space. In our case, at least concerning the experiments,  the fractal dimension $D_2$ is measured 
on two dimensional cuts of the flame front surface. 
Assuming isotropic properties of the flame front,  the dimension of the whole 3D surface is expressed as: $ D = D_2 +1$.
In real surfaces, the fractal scaling characterised by the dimension $D$ 
usually hold only in a limited range of scales between an inner ($\epsilon_i$) and an outer ($\epsilon_o$) 
cut-off lengths.
Fractal characteristics of our experimental and numerical dataset 
have been evaluated by the box-counting technique, which
consists in enumerating the squared boxes of size $L$ necessary to
entirely cover the whole object that, in this case, is the flame
front. If this number scales with equation \eqref{scal_D}, then the measured
object is a fractal with dimension $D$ (or $D_2$ in 2D cuts).
Right panel of figure \ref{fig:4a} reports the fractal scaling  of two flames,
 one measured and the other simulated. 
The plots show nice scalings maintained up to the inner cut-off length $\epsilon_i$, whose value
has a direct impact on the wrinkling factor of equation \eqref{eq:1b}; 
outer cut-offs are also apparent, though less important for LES applications. 
Data at different $u'/S_L$ have been fitted with a power law whose exponents are 
reported in  the left panel of figure \ref{fig:1}. The fractal 
dimension $D$ shows an almost constant value of $D = 2.23 \pm 0.03$, well lower than $D = 2.37$ which
corresponds to perfectly passive scalar iso-surface fractal dimension
as evaluated by \cite{sreenivasan1986fractal}. 
It should be remarked that, for the simulation dataset, the fractal dimension $D$ directly 
obtained by the three-dimensional box counting method on the whole flame front ($c=0.5$) and those extracted from
the 2D cuts $D=D_2+1$ do not show significant differences and are in very good agreement with
the experimental values. 
Hence, for the different flame regimes here analysed, from corrugated to the beginning of thickened flames, 
the fractal dimension $D$ appears to be
a very stable quantity that does not depend  either on $u'/S_L$ or on other parameters, see also~\cite{gulder1}. 
%
Concerning the
inner cut-off length $\epsilon_i$, two different scaling are supposed to exist: It may scale either with the Kolmogorov dissipative length 
$\eta_k$ or with the laminar flame thickness $\delta_L$. Normalising the inner cut-off length by the laminar flame thickness 
$\delta_L$, it holds: 
\begin{equation}
{\epsilon_i} / {\delta_L} \propto  Ka^{\beta},
\label{scal_Ka}
\end{equation}
where $\beta=0$ indicates $\epsilon_i\propto \delta_L$, while $\beta=-1/2$ denotes the scaling of the inner cut-off proportional 
to the Kolmogorov length. In the right panel of figure \ref{fig:1} we report the scaling of ${\epsilon_i} / {\delta_L}$ for the experimental
dataset. The scaling with $\beta=-1/2$ appears to best fit the data. In particular, the curve $Ka^{-1/2}$
almost perfectly correlates to homogeneous data denoted by the same symbol. These groups correspond to the same experimental conditions concerning the turbulence features $\eta$, $u'$ and $Re$, 
but differ for the equivalence ratio which controls $S_L$ and $\delta_L$.  
{Hence, even with $Ka=(\delta_L/\eta)^2>1$, we found a 
cut-off length proportional to the Kolmogorov scale $\eta$, i.e.\ $\epsilon_i\propto \eta$, but
showing values  of the order or larger than the corresponding flame thickness, i.e.\ $\epsilon_i\ge\delta_L$. 
This is cleared in the inset
of the right panel of figure~\ref{fig:1} where the inner cut-off rescaled by the Kolmogorov length $\epsilon_i/\eta$ appears
almost constant for all the cases with a value around $\epsilon_i/\eta\simeq10$.\\
This peculiar behaviour can be explained considering that the fractal  wrinkling of the flame front is produced by the
turbulent self-similar eddies characterising the Kolmogorov inertial range (\emph{-5/3 law}). Hence it is expected that the fractal
features of the flame wrinkling are lost at the end of the inertial range.
As shown by the data 
reported in~\cite{sadvee} (see fig.~9 of the paper) the Kolmogorov's universal scaling of the inertial range (\emph{-5/3 law}) is 
actually lost at a scale $\ell$, here referred as `limit scale', proportional to the Kolmogorov length $\eta$, although 
significantly larger.
Specifically, from the power spectra reported in~\cite{sadvee} we can estimate this `limit scale' as
$2\pi\eta/\ell=0.3\div0.6$, i.e.\ $\ell\simeq10\div20\eta$, meaning that the self-similar inertial range is actually 
lost at about $10\div20$ Kolmogorov length. 
Concerning our dataset, we found that up to the highest Karlovitz number here investigated, i.e.\ $Ka=150$, the inner cut-off
of the fractal flame front is lost when $\epsilon_i\simeq10\eta\simeq\ell$. 
It should be noted, that the present dataset cannot shed light on what happens at very high $Ka$ where the typical velocity structures of the inertial range 
 can directly interact with the flame front. It is known that increasing even more the Karlovitz number, i.e.\ $Ka\gg100$, 
 the local structure of the flame front becomes strongly altered hence thickened, broken or distributed turbulent flames take 
 place~\cite{peters2000turbulent}. 
 Since the present modelling relies on a local flame structure similar to the laminar flame (flamelet approximation), we guess that in the 
very high $Ka$ regime the present framework cannot be directly applied. 
}

\section{Large-Eddy-Simulations}
The results previously discussed on the fractal dimension $D$ and on the
inner cut-off $\epsilon_i$ will be exploited to perform LES of reactive flows in order
to determine the a-posteriori performance of this simple and fast model.  

\subsection{LES algorithm}

The same numerical algorithm previously used for DNS  
solves the Favre-filtered Low-Mach Navier-Stokes equations,
%
\begin{align}
\label{eq:cont_def_les}
\frac{\partial \bar{\rho}}{\partial t}& + 
\nabla \cdot \left( \bar{\rho} \tilde{\vec{u}} \right)=0,
\\
\label{eq:mom_def_les}
\frac{\partial \bar{\rho} \tilde{\vec{u}}}{\partial t} &+ 
\nabla \cdotp \left( \bar{\rho} \tilde{\vec{u}} \otimes \tilde{\vec{u}} \right) 
= \nabla \cdot {\vec R}_{sgs}
- \nabla \bar P +\frac{1}{Re} \dive \overline{ \Sigma}+\bar{\rho} {\vec{g}},
\\
\label{eq:prog_var_les}
\frac{\partial \overline{\rho}\tilde c}{\partial t} &+ {\bf \nabla} \cdot (\overline{\rho}\tilde {\bf{u}}\tilde c)=
\nabla \cdot {\vec D}_{sgs} +\rho_u S_{L} \Xi \vert\nabla \tilde{c}\vert
%
\\
\label{eq:div_def}
\nabla \cdot \tilde{\bf u} &=
\frac{1}{p_0 } \left\{ \nabla \cdot {\vec T}_{sgs} + \frac{\gamma-1}{\gamma} Ce \rho_u S_L \Xi \, |\nabla \tilde{c}|\right.\\
&+ \left. \frac{1}{Re}\left(\frac{1}{Pr } \nabla \cdot ( \overline{k} \nabla \tilde{T}) 
-\frac{Ce}{Sc} \frac{\gamma-1}{\gamma} \nabla \cdot ( \overline{\rho {\cal D}} \nabla \tilde{c})\right)
 \right\}\\
\label{eq:state_def}
& \overline\rho=\frac{p_0}{1+ Ce \frac{ \gamma-1}{\gamma}\, \tilde{c} }
\end{align}
where we assumed $Pr=Sc=0.7$ ($Le=1$).
The following extra-stress term is originated by applying the filter to the momentum equation~\eqref{eq:mom_def}: 
\begin{equation}
{\vec R}_{sgs}=-  \bar \rho \left( \widetilde{ \vec{u} \otimes \vec{u}}-
\tilde{ \vec{u}} \otimes \tilde{\vec{u}}\right)= - 2 \overline\rho \vec I 
\left( \frac{k_{sgs}}{3} + \dive \tilde{\vec u}\right) 
+2\, \overline\rho \nu_{T}\,\widetilde{\vec S} ,
\label{eq:fsgs}
\end{equation}
with $2 k_{sgs}=\widetilde{|{\vec u}|^2}-|\tilde{\vec u}^2|$. It 
was modelled by using  the Smagorinsky model with shear-improved formulation by \cite{shear-impr},
rearranged to take into account the variations of density. 
%
$\tilde{S}= 0.5 \left(\nabla \tilde{\vec u} + (\nabla \tilde{\vec u})^T\right)$ is the resolved rate of strain 
tensor and 
$\nu_T= (C_S \Delta)^2 \cdot (| \tilde{S}({\vec x},t)|- | \langle\tilde{S}({\vec x},t)\rangle|)$  is 
the  eddy-viscosity with the constant $C_S = 0.12$ and the typical cell size $\Delta=\sqrt[3]{r \Delta\theta\Delta r \Delta z}$, 
for details see~\cite{shear-impr}.
Sub-grid extra-diffusion terms of scalar quantities read,
\begin{align}
{\vec D}_{sgs}=- \bar{\rho} \left( \widetilde{{\vec u}\,c} - \tilde{\vec u} \tilde{c}\right)= 
- \bar{\rho}D_T \nabla \tilde{c}\\
{\vec T}_{sgs}=- \bar{\rho} \left( \widetilde{{\vec u}\,T} - \tilde{\vec u} \tilde{T}\right)= 
- \bar{\rho}\alpha_T \nabla \tilde{T}
\end{align}
and the sub-grid diffusivity is assumed proportional
to the Smagorinsky eddy-viscosity: $D_T=\alpha_T=0.7 \nu_T$, constant
turbulent Schmidt and Prandtl numbers.
The wrinkling factor in the \eqref{eq:prog_var_les} is modelled using the fractal features here
extracted by the numerical and experimental investigation,   
$$\Xi(\Delta)=\alpha(\Delta/\epsilon_i)^{D-2},$$ 
 where $\alpha=1.7$ is assumed together with 
$D=2.23$ and $\epsilon_i= 10 \eta = 10(\nu^3/\varepsilon)^{1/4}$ 
 with $\varepsilon$ the turbulent kinetic energy dissipation.
The Kolmogorov scale $\eta$ is
calculated using filter observables and assuming K41 theory. If the LES filter scale 
$\Delta$ falls inside the inertial range, the dissipation can be 
estimated as $\varepsilon\simeq  \tilde u_\Delta^3/\Delta$, where
the characteristic velocity at scale $\Delta$ 
is  $ \tilde u_\Delta= \tilde S \Delta$ with $\tilde S$ the 
filtered strain rate. 

Several Large Eddy Simulations {concerning premixed round jets at different Reynolds
numbers and $U_b/S_L$ have been performed in order to test the model in different conditions.
Firstly, an accurate comparison between DNS and LES data at the
 Reynolds number $Re=6000$ and corresponding numerical conditions will be
presented.
The mesh of the LES is obtained reducing by a factor $4$ in each direction 
the grid points of the corresponding DNS mesh, $N_\theta\times N_r\times 
N_z=32\times51\times140$.\\
Successively, keeping fixed the model parameters, two LES' 
at higher Reynolds numbers and $U_b/S_L$, i.e. $u'/S_L$, have been performed 
in order test the model at conditions close to real applications.
In this case LES' reproduce the experimental configurations of~\cite{cheman} with  
$Re=U_b D/\nu=16000$ (M3) and $Re=24000$ (M2), respectively. 
Computational domain and mesh coincide
in the two cases and are given by:
$[\theta_{max}\times R_{max} \times Z_{max}]=[2\pi \times 6.2D \times
17.5D]$ and $N_\theta\times N_r\times N_z=64\times101\times700$, respectively.
Similarly to the DNS procedure, fully turbulent inflow conditions are  
provided by a companion LES of a periodic turbulent pipe flow at the same Reynolds number.
The adopted resolution is estimated to be 80 and 180 times smaller than what needed for a resolved DNS. However 
 it is expected to be sufficient to well simulate incompressible
jets even at higher Reynolds number, i.e. $Re=400000$, where the minimum requirement for mesh size has been found to be 
the shear length $L_s=\sqrt{\varepsilon/{\cal S}^3}$ (where ${\cal S}$ is the shear-rate modulus), see~\cite{pichan_ftc12} for
more details on LES of incompressible jets and on their resolution issue. These aspects assure that the turbulence model
should properly simulate the turbulent jet dynamics and allow to test the LES performance for combustion.

\subsection{LES results}

\begin{figure}[t]
\centering
\includegraphics[width= .25\textwidth]{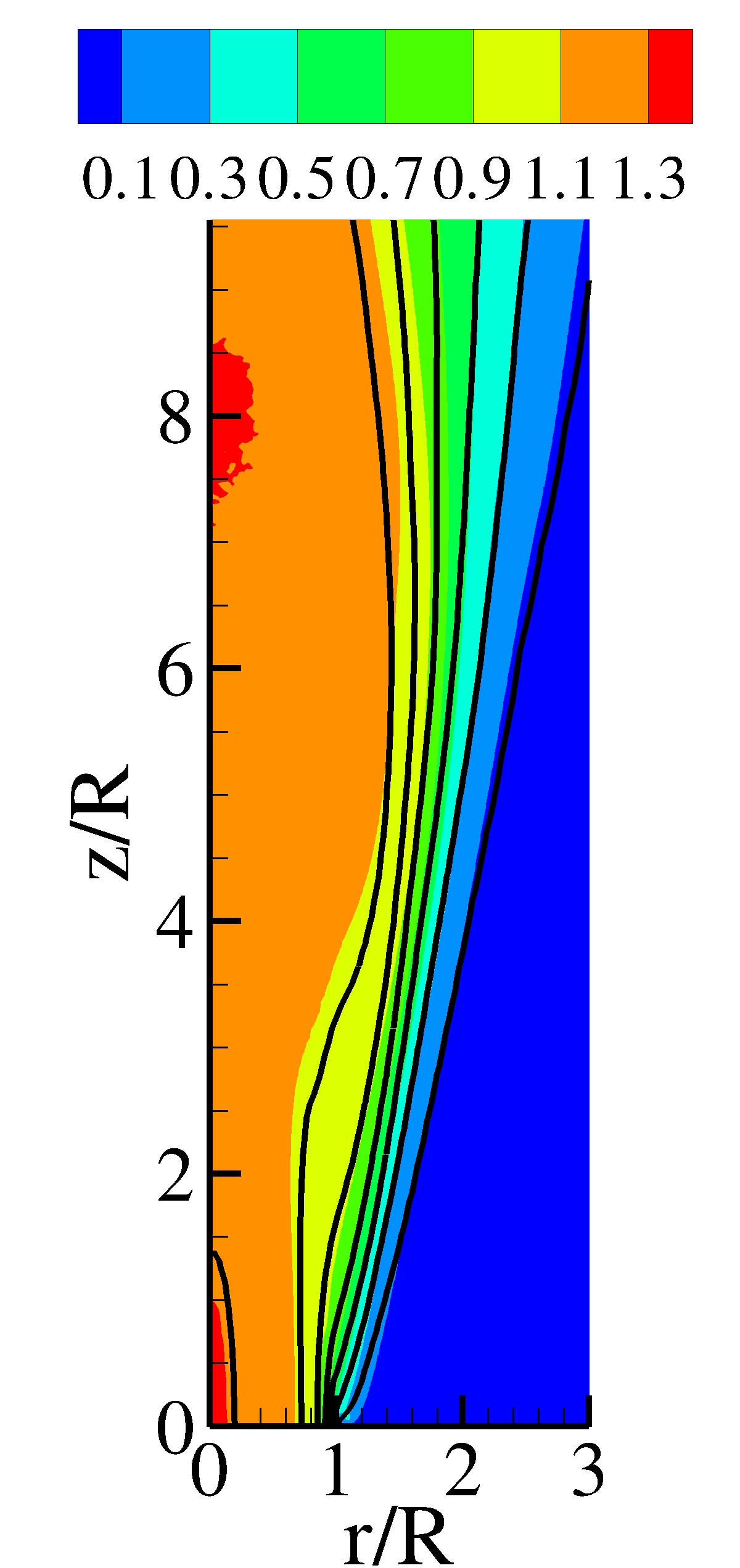}\hspace*{.4cm}
{\put(-72,210){$U_z/U_b$}}
\includegraphics[width= .25\textwidth]{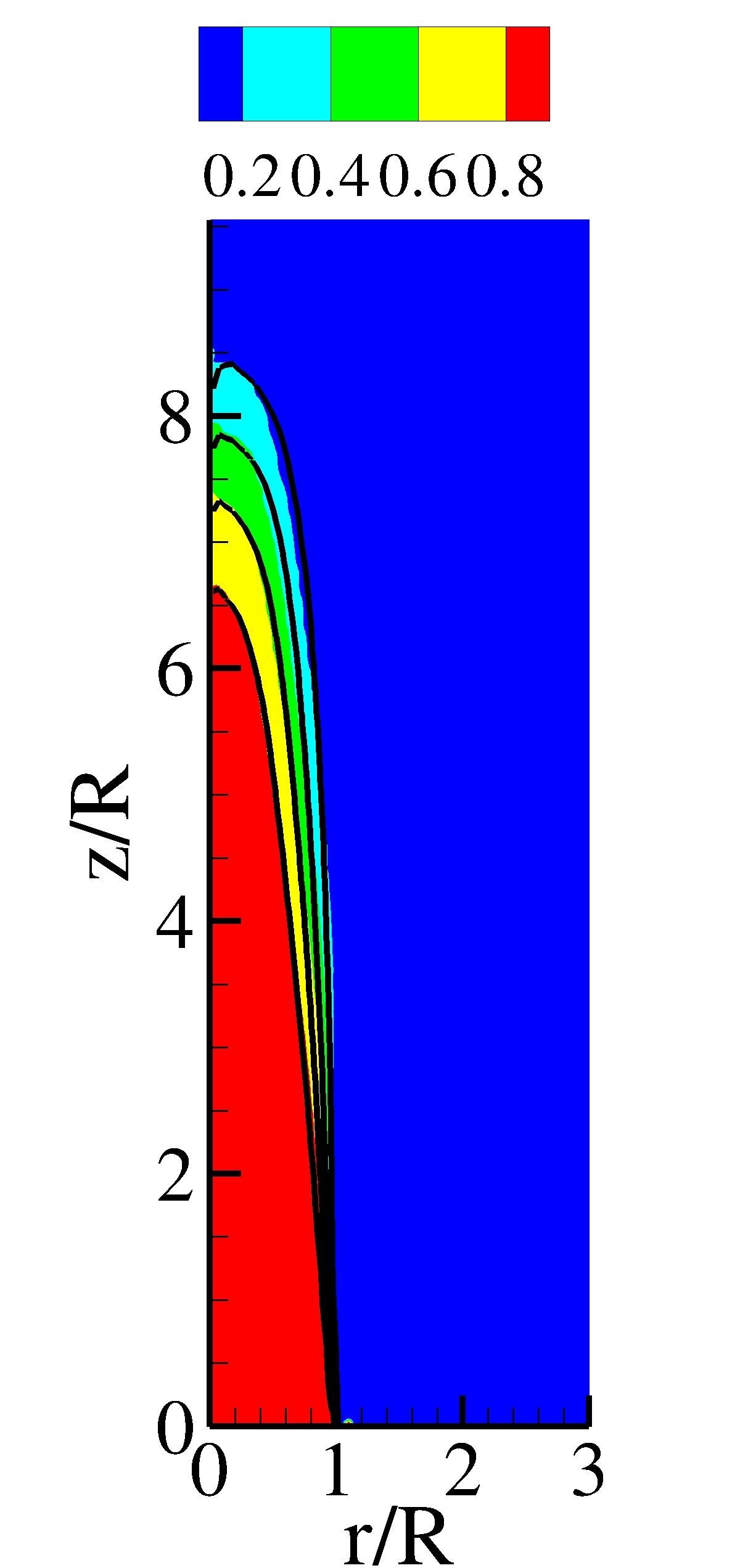}\hspace*{.2cm}
{\put(-70,210){$Y_R/Y_R^0$}}
\includegraphics[width= .25\textwidth]{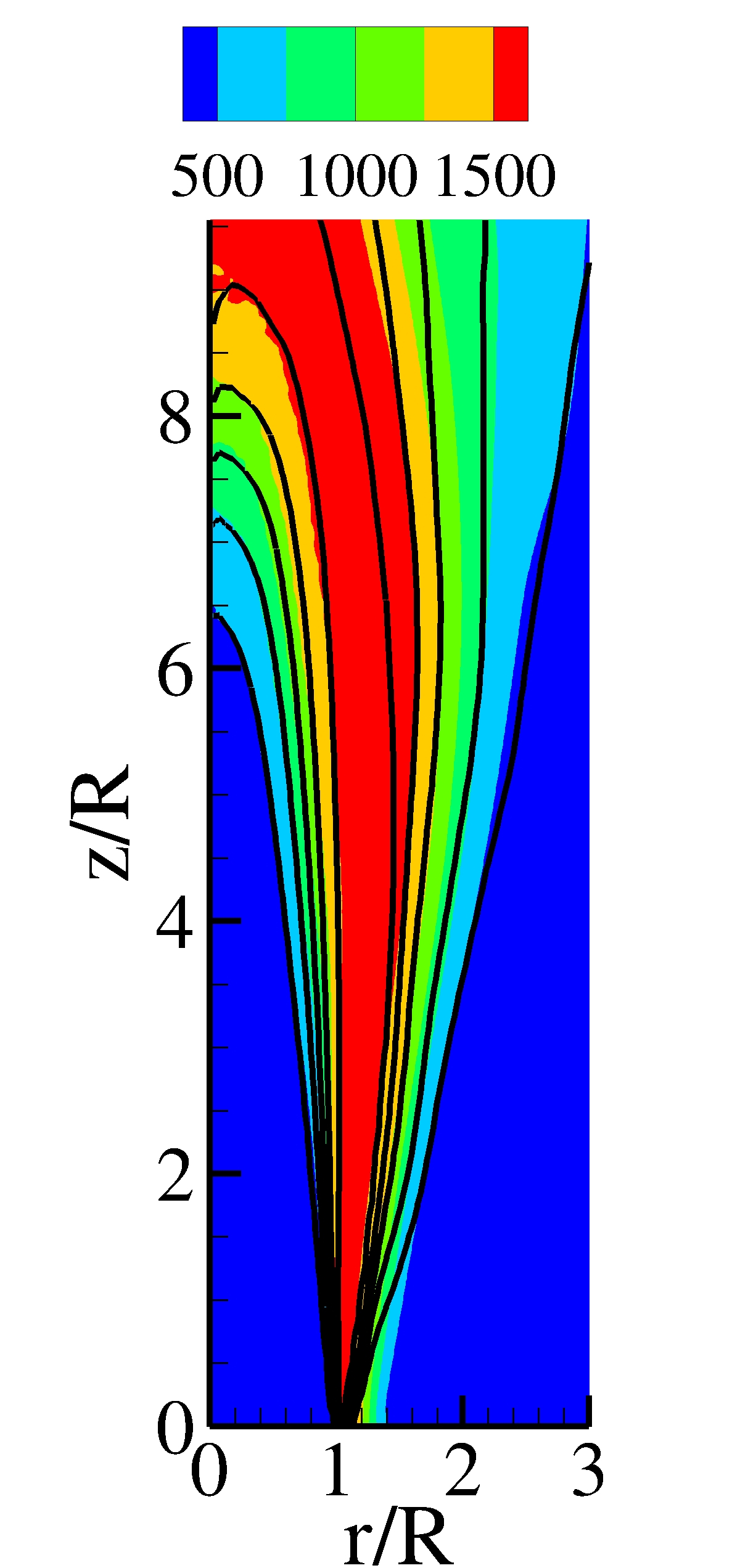}
{\put(-50,210){$T$}}
\caption{\label{fig:field_comp} Favre-averaged mean fields of axial velocity $U_z/U_b$ (left panel), 
$Y_R/Y_R^0$ (middle panel) and dimensional T (right panel). Contours, DNS with $Re=6000$ and
$u'/S_L=2$; Thick black lines corresponding LES with mesh 4 times coarser in each direction.  
 }
\end{figure}

A detailed analysis of the LES model performance 
is carried out replicating the corresponding configuration of the DNS of
the premixed Bunsen flame at $Re=6000$ and $u'/S_L=2$ with a mesh $4^3$ times coarser.
A general overview of the mean fields is provided in figure~\ref{fig:field_comp} where
contours represent DNS data and thick black lines the corresponding levels from  LES.
In the velocity field, the effect of the expansion due to the heat release induces a spread
of the mean axial velocity not associated with the typical decay of incompressible jets~\cite{pic_cas,pichan_ftc12}.
The LES model appears to well reproduce this behaviour implying a correct representation of the heat release.
In the middle panel the reactant concentration has been presented. The prediction of the flame brush $0<Y_R/Y_R^0<1$, 
i.e. the region spanned by the instantaneous flame front, is quite remarkable since both the flame height and width are precisely 
estimated by the LES model. It means that the wrinkling factor $\Xi$ is correctly approximated by the fractal model with $D=2.23$
and $\epsilon_i\propto \eta$.  The right panel of figure~\ref{fig:field_comp}, shows the dimensional temperature field 
($T_0=300\,K$). It should be noted that this Bunsen flame reacts in an open environment, so the temperature iso-levels coincides
with the reactant concentration iso-levels only in the inner part of the flame brush where the mixing with the cold environment is
{negligible}. However, further away from the maximum temperature that almost coincides with the 
adiabatic one, the mixing with the ambient air (entrainment) occurs leading to an additional mixing layer not present in the 
reactant concentration. As apparent from 
the plot, the flame brush is correctly captured by the LES model, while small differences are present in the outer mixing layer.

To better analyze the performance of the LES in reproducing DNS data, we report in figure~\ref{fig:4} the radial profiles of the same 
Favre-averaged quantities at four axial stations, $z/D=1;\,2;\,3;\,4$, for DNS and LES. The mean axial velocity
$U_z/U_b$, left column, shows an excellent matching in the
whole domain.
Profiles of the mean reactant concentration $Y_R/Y_R^0$ --middle column of
figure~\ref{fig:4}-- also shows a good agreement. As can be appreciated in figure~\ref{fig:field_comp}, 
further downstream, e.g. at $z/D=5$, the mean concentration of reactant is vanishing
 both for DNS and LES indicating that the flame brush and height are accurately captured by the LES model.

\begin{figure}
\centering
\includegraphics[width= .327 \textwidth]{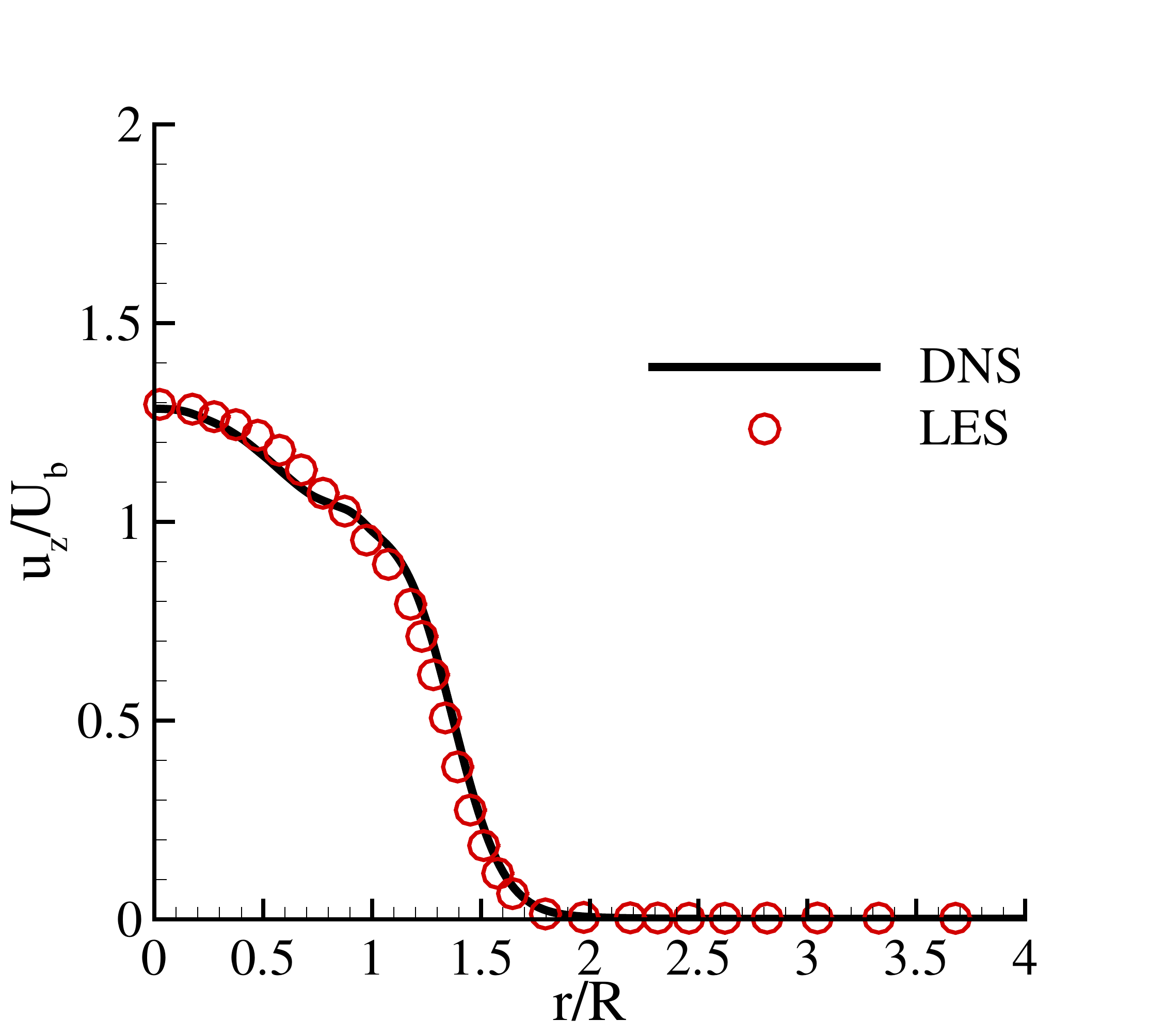}
\includegraphics[width= .327 \textwidth]{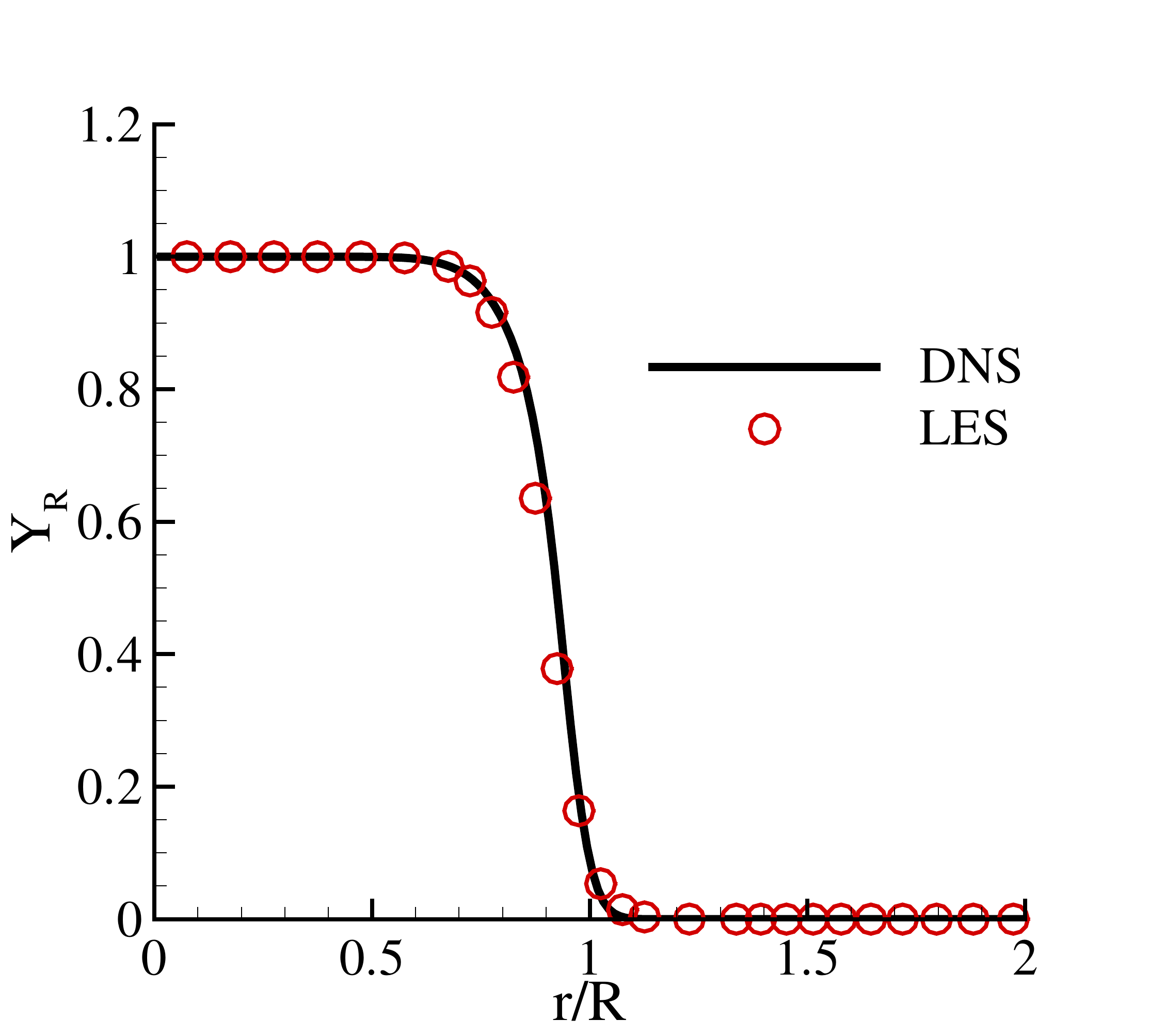}
\includegraphics[width= .327 \textwidth]{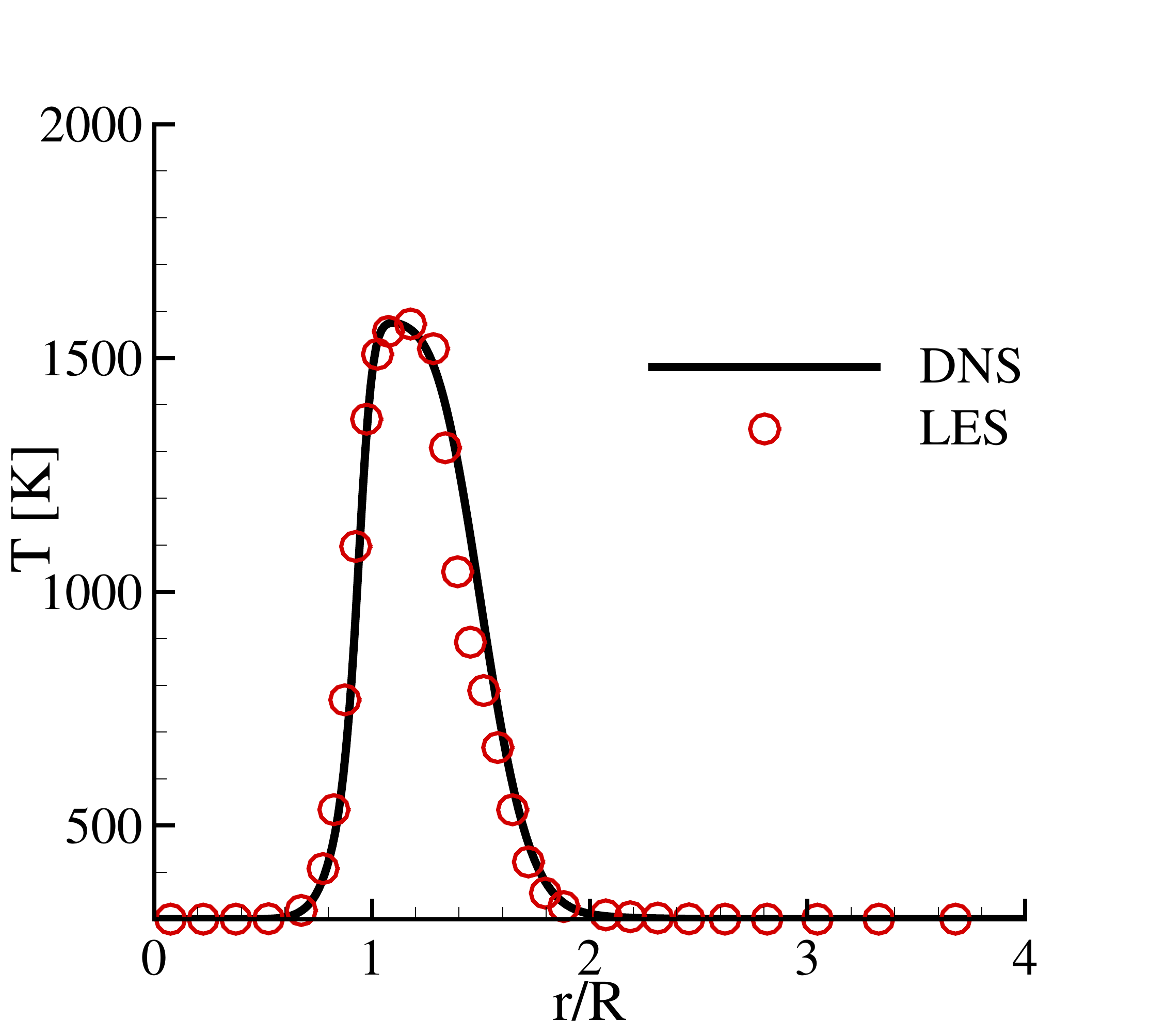}\\[.2cm]
\includegraphics[width= .327 \textwidth]{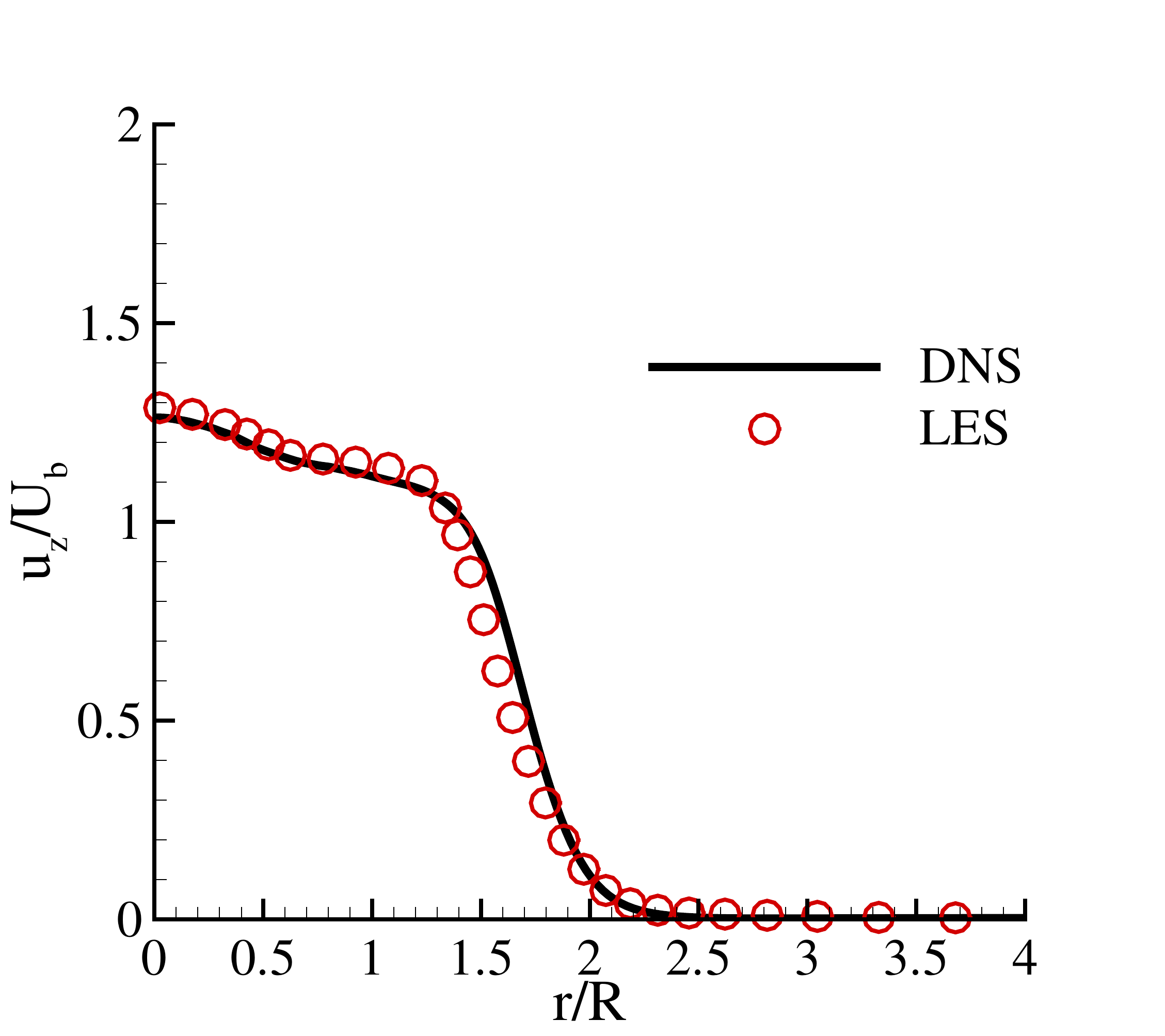}
\includegraphics[width= .327 \textwidth]{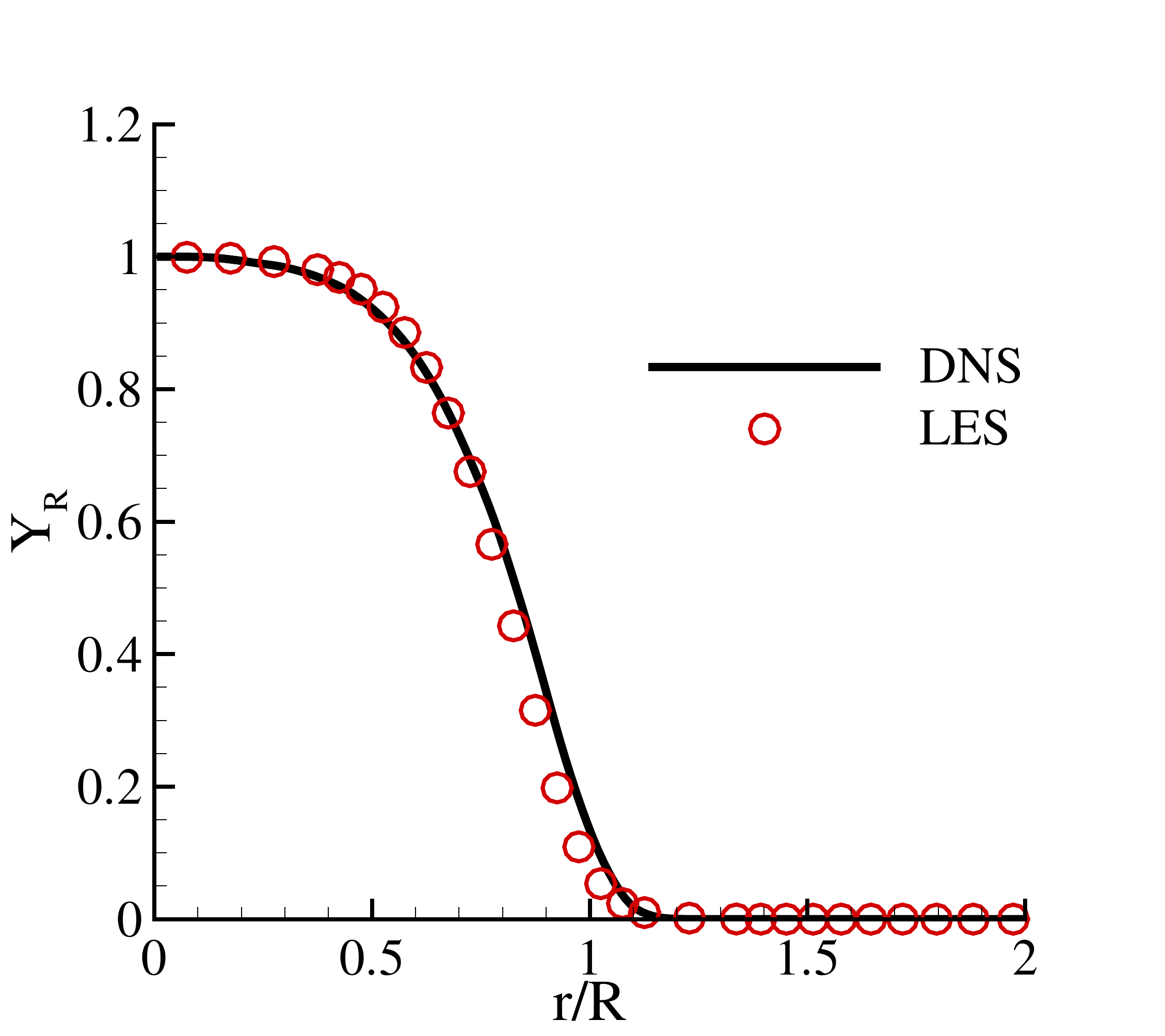}
\includegraphics[width= .327 \textwidth]{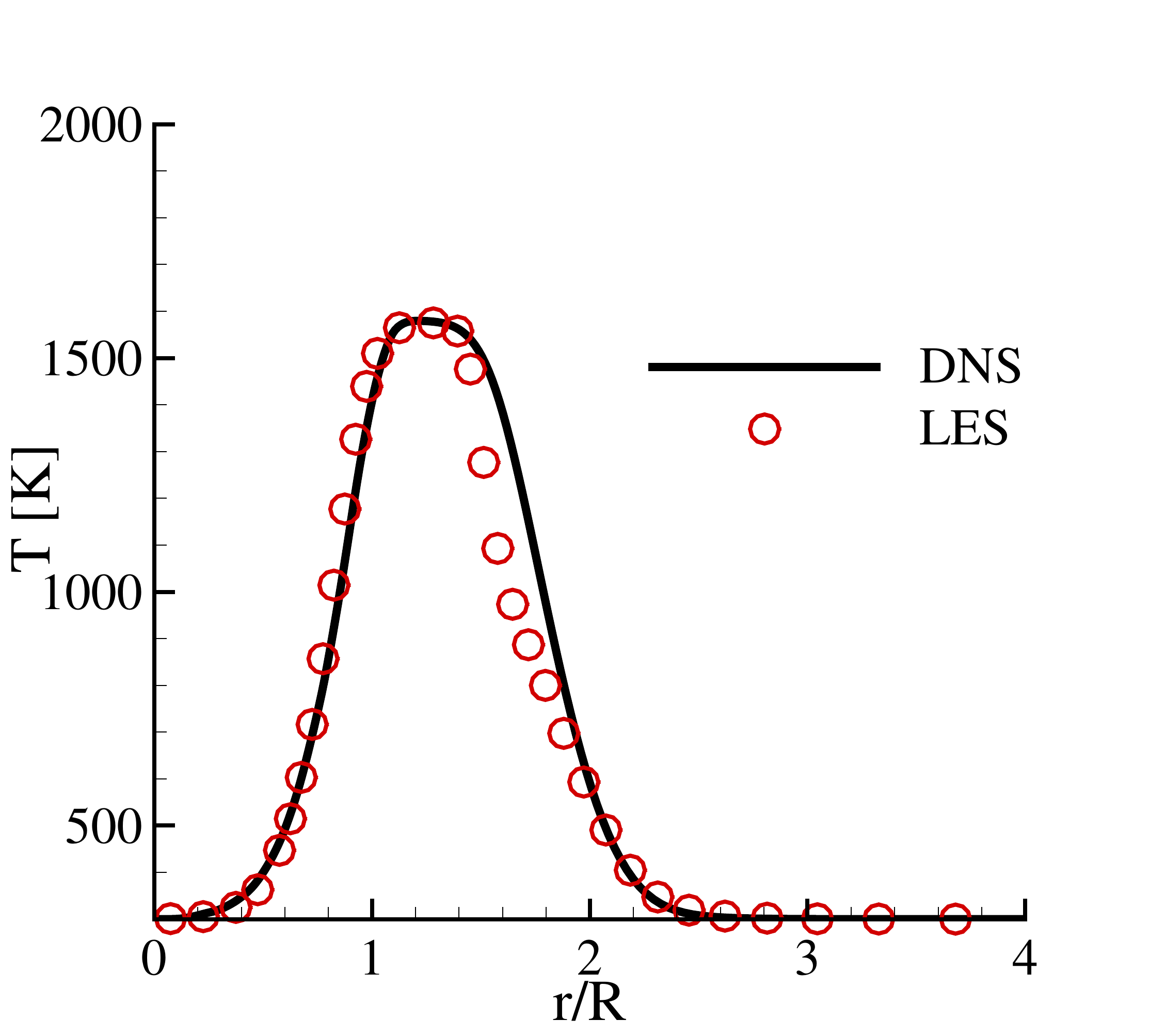}\\[.2cm]
\includegraphics[width= .327 \textwidth]{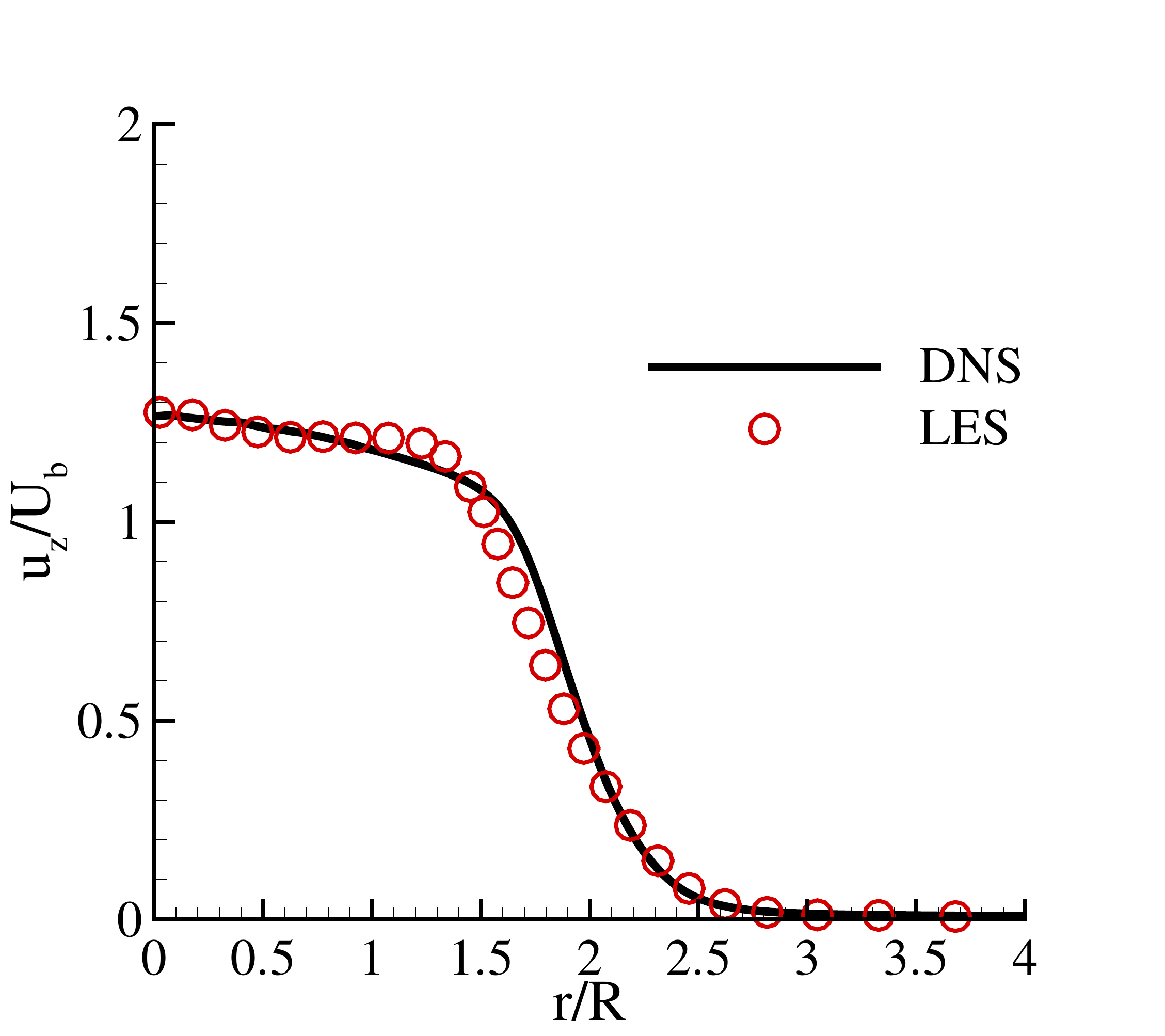}
\includegraphics[width= .327 \textwidth]{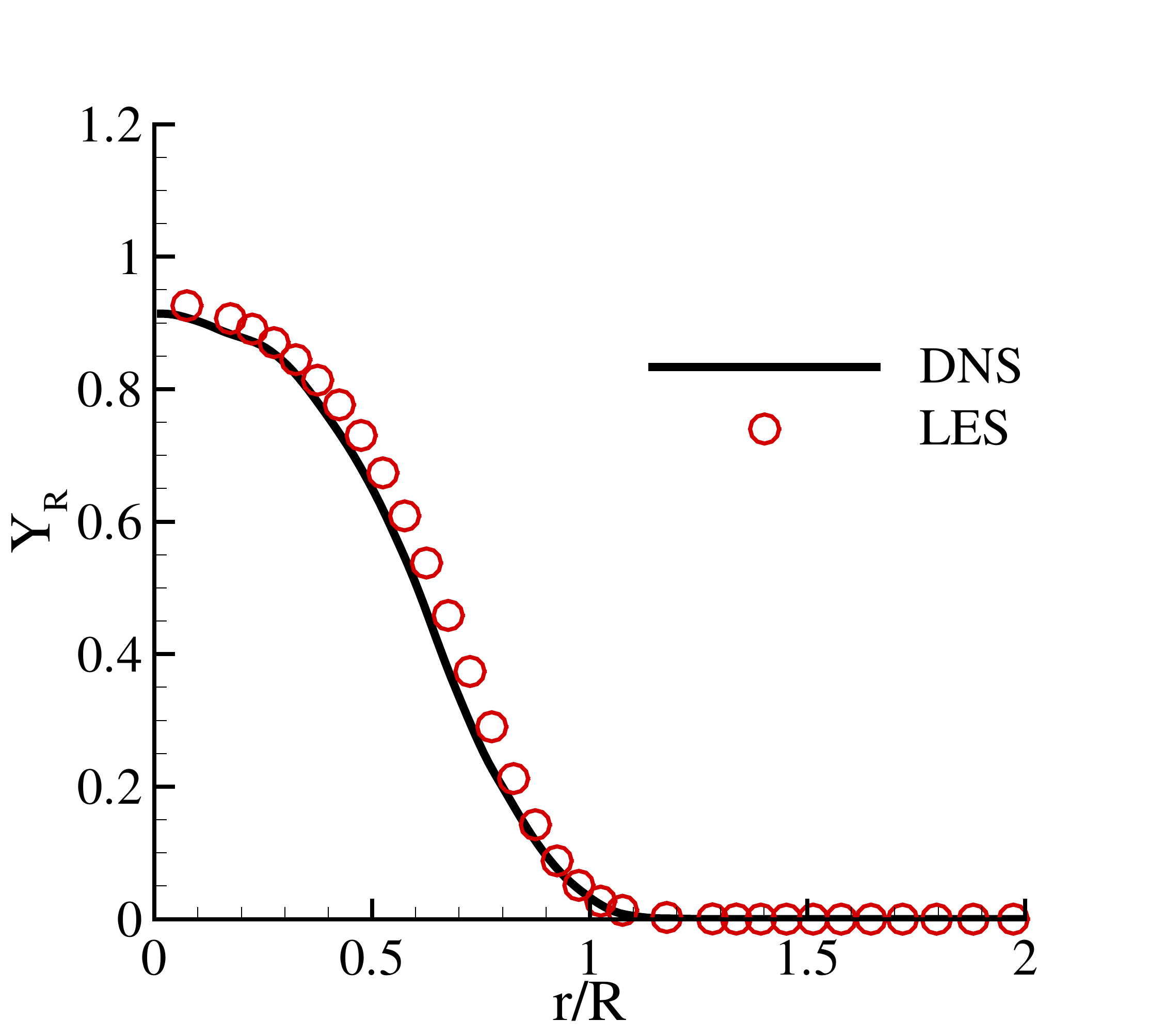}
\includegraphics[width= .327 \textwidth]{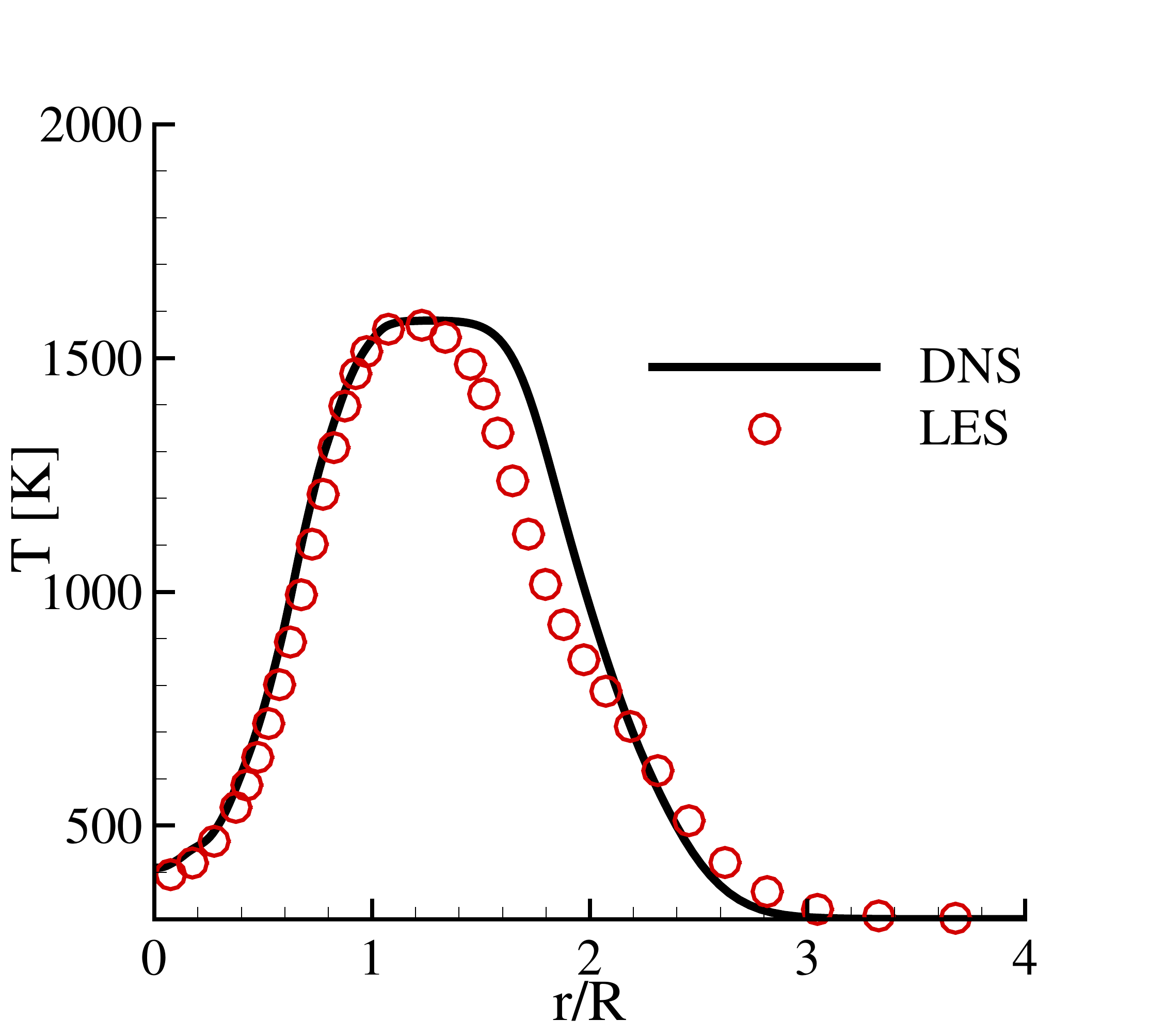}\\[.2cm]
\includegraphics[width= .327 \textwidth]{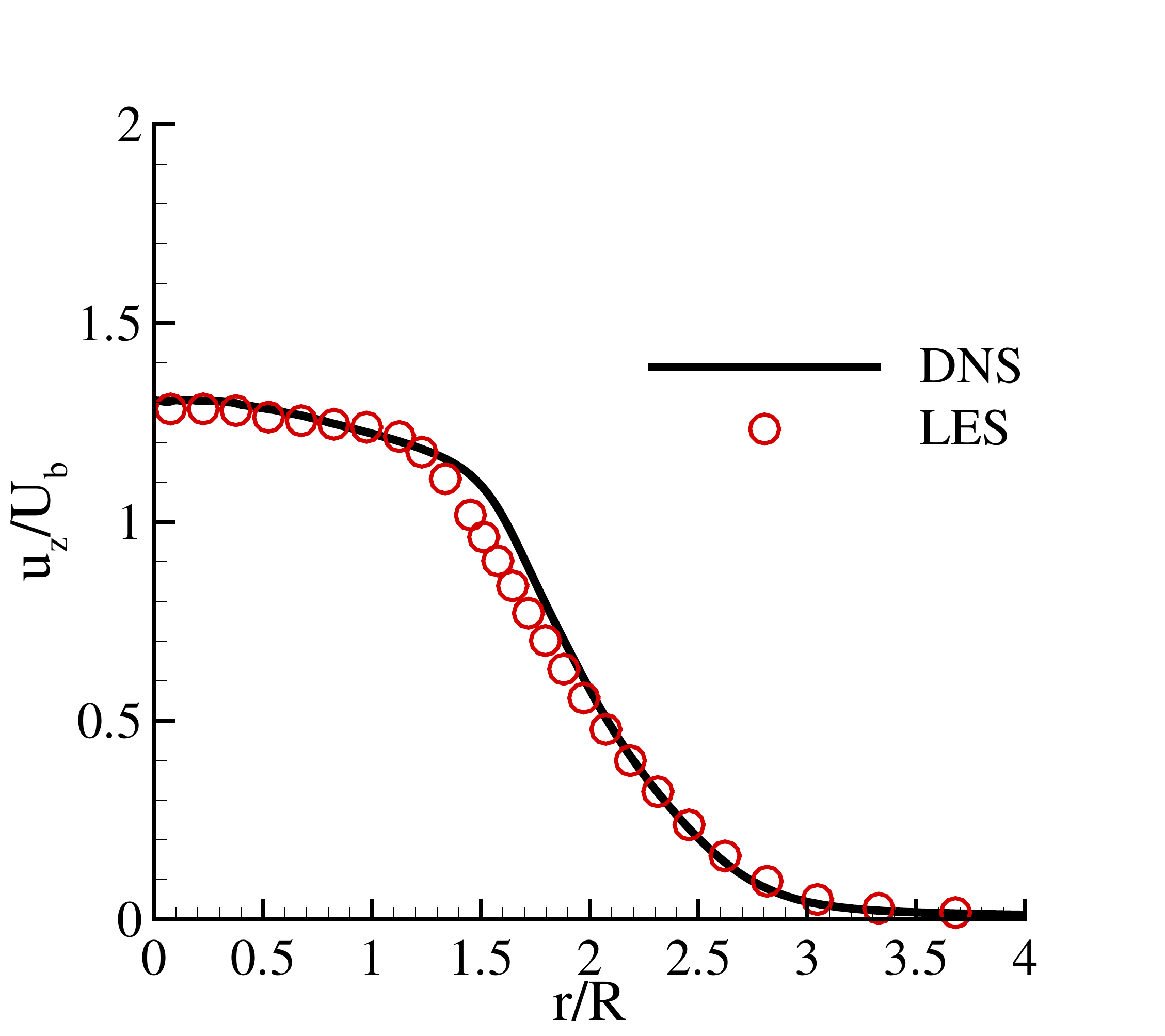}
\includegraphics[width= .327 \textwidth]{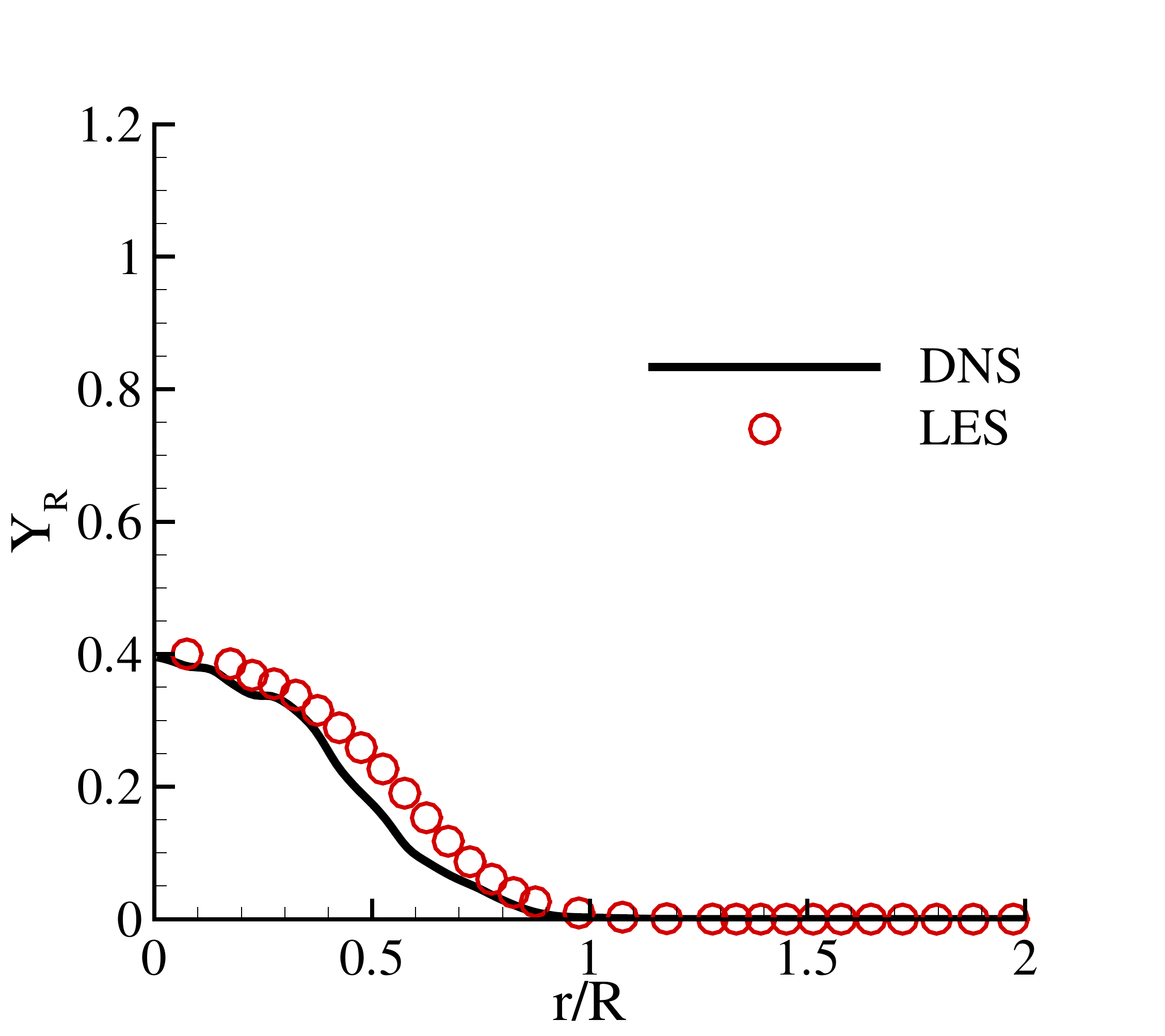}
\includegraphics[width= .327 \textwidth]{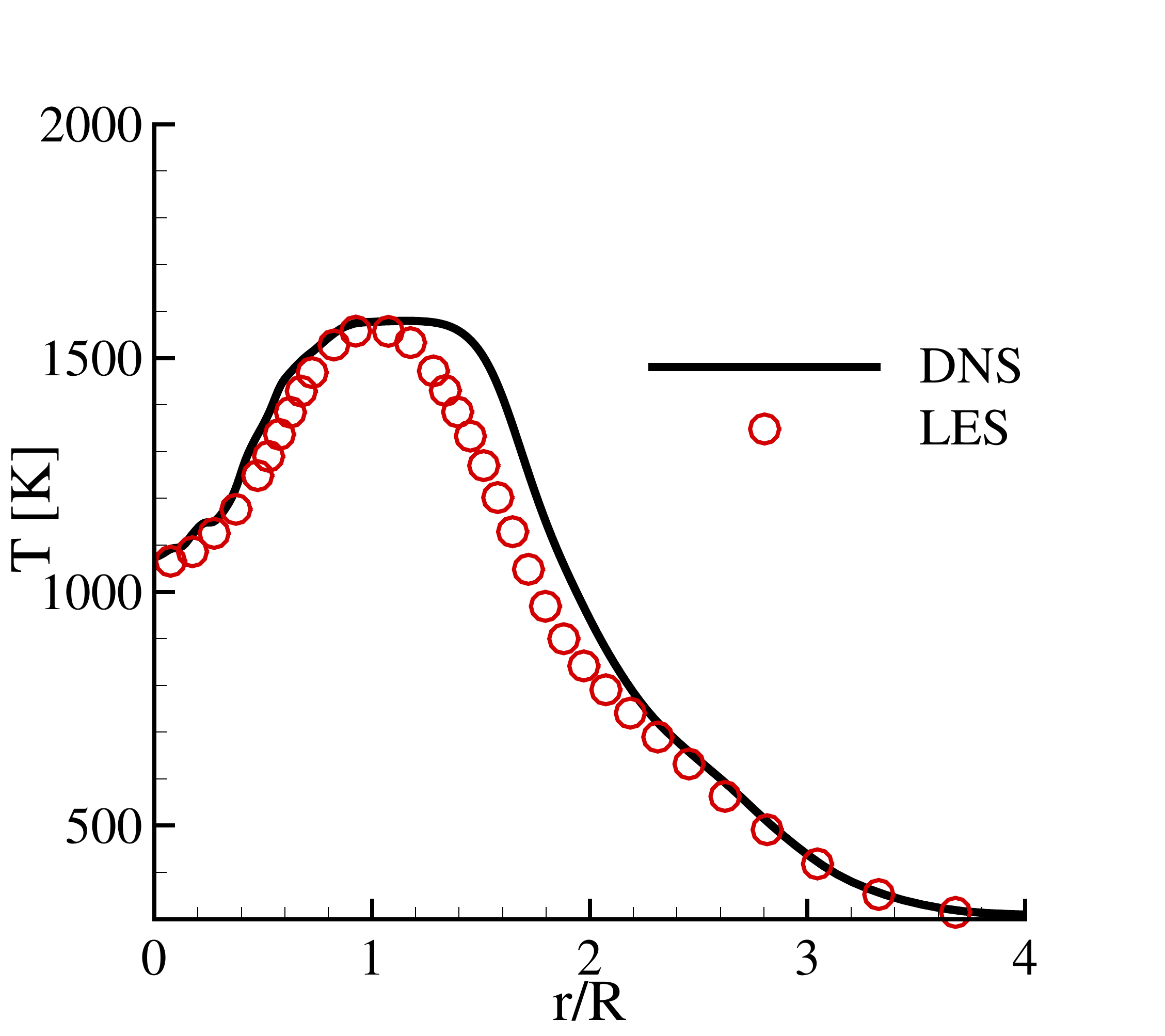}
\caption{ Comparison between LES and DNS at $Re=6000$ and $u'/S_L=2$.
Left column: Favre-averaged axial velocity profiles $ U_z/U_b$; 
middle column: Favre-averaged reactant profiles normalised by inlet concentration 
of reactants $Y_R/Y_R^0=1-C/C^0$; right column: mean temperature profiles $T$ (dimensional).
Rows corresponds to sections at four axial distances, from top: 
$z/D=1$,
$z/D=2$,
$z/D=3$,
$z/D=4$.  }
\label{fig:4}       
\end{figure}
Right-most panels of figure~\ref{fig:4} report the mean temperature profiles $T$ (dimensional with $T_0=300K$). 
In each section the temperature shows a maximum almost 
matching the adiabatic flame temperature ($T_f\simeq 1550\,K$)  in a region where only burnt gases are present, 
while for larger radial distances the ambient temperature is approached. 
Also for this observable, we find a good accordance with DNS data, 
with small differences present in the mixing region between hot gases and
surrounding air where  chemical reactions are not present anymore. 
Actually, the mixing properties in this region show peculiar features that
 turbulence models often fail to correctly reproduce~\cite{da2009behavior}. 
In this zone the entrainment of ambient air takes place across a fluctuating intermittent layer which
separates the turbulent jet core and the irrotational environment. The non fully turbulent behaviour may be responsible
for the  deviations we observe in this region.

\begin{figure}
\centering
\includegraphics[width= .3 \textwidth]{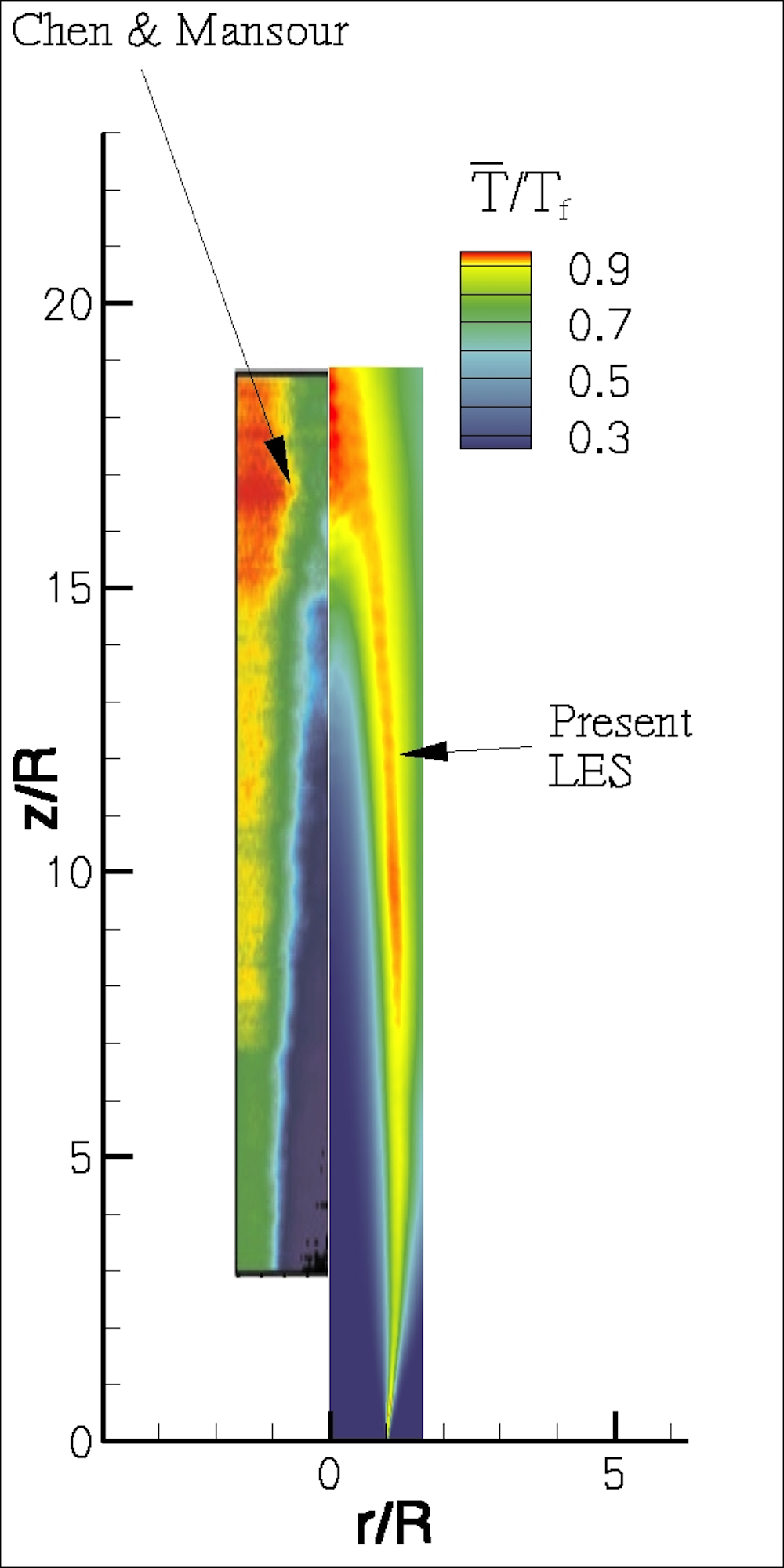}
\hspace{1cm}
\includegraphics[width= .3 \textwidth]{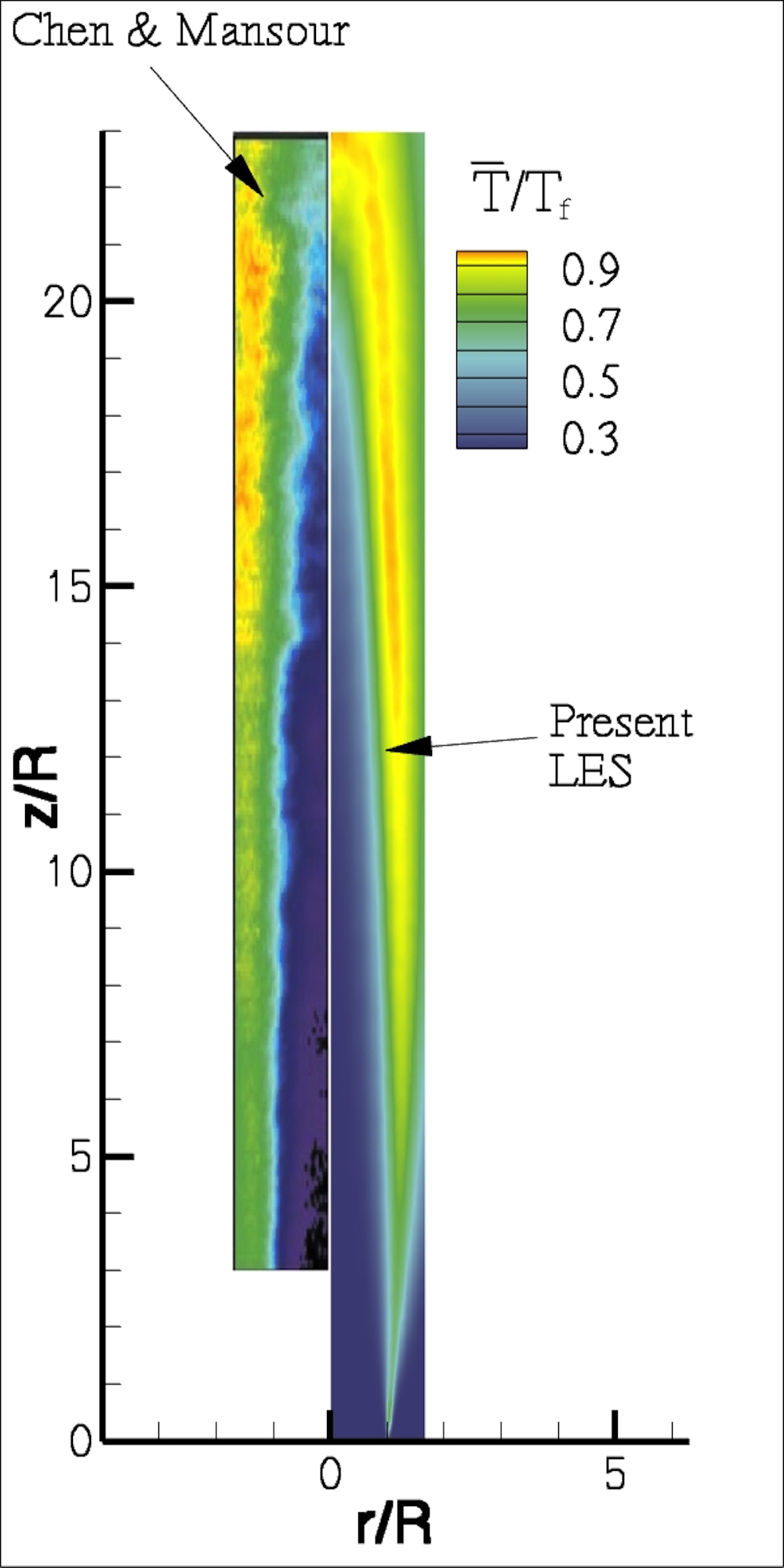}
\caption{Mean temperature field normalised with the adiabatic flame temperature.
Left half-figure: experiments of {\it Chen and Mansour}~\cite{cheman}.
Right half-figure: present LES computations.
Right panel, $Re = 16000$, $S_L/U_b=0.019$; 
Left panel, $Re = 24000$, $S_L/U_0=0.013$.
\label{fig:7}       
}
\end{figure}

To assess the behaviour of the LES at higher Reynolds number we have simulated two  turbulent premixed jet flames
at $Re=24000$ and $Re=16000$, whose  data have been taken from the
experiments of 
\cite{cheman} on $CH_4/Air$ flames (M2 and M3). 
The ratios between laminar unstretched flame speed and bulk jet velocity are
$S_L/U_b=0.013$ and $S_L/U_b=0.019$ for the two flames, respectively.\\
In figure~\ref{fig:7}, the mean temperature field normalised by the adiabatic
flame temperature $T/T_f$ is presented (left mid-figures) and compared to that obtained experimentally (right mid-figures).
The  flame height and width appear correctly estimated by LES. It should be considered that 
model parameters and mesh size have been kept fixed for the two cases indicating that 
the present model is sufficiently stable to change of physical conditions.
We remark that the present LES mesh is about $80$ and $180$ times smaller than that needed for a resolved DNS 
at $Re=16000$ and $Re=24000$, respectively.
Some differences between experiments and LES are present in the lower part of the shear
layer and especially at the flame tip, where the experimental temperatures are lower than
those estimated by LES. 
We attribute these phenomena to local flame stretch and curvature effects induced by the  high shear 
rate, which may modify flame propagation and even lead to local flame extinction, i.e.\ quenching. 
This issue has been discussed in details in the work~\cite{cheman} and attributed to the
 high shear rate typical of these high-Reynolds number flames.
{Actually, the present LES does not incorporate any `sub-model' able to reproduce the quenching due 
to a high local value of the shear rate. In the present LES model, a local high shear rate induces a small 
local Kolmogorov length and hence a high wrinkling factor $\Xi$.} 
It should be noted that it is straightforward to incorporate
directly in $\Xi$ a model for the local quenching, e.g.\ $\Xi=0$ if the local shear rate exceeds a threshold. The 
issue is out of the scope of the present work, being still under investigation by the present authors. 

\section{Final remarks}

In the present study we investigate the fractal characteristics of turbulent premixed jet flames 
via a combined approach using  DNS and laboratory experiments 
{up to a Karlovitz number $Ka=150$}.  Experiments on Methane/Air flames consider different 
inflow geometries, i.e.\ annular and round Bunsen jets, Reynolds numbers and equivalence ratios. 
Concerning numerical simulations, a Methane/Air lean mixture has been 
reproduced by a simple one-step reaction reproducing the fundamental 
behaviour of a round Bunsen flame. The main flame characteristics as well as its multi-scale fractal structure
have been found in 
good accordance with the experiments. 

The fractal dimensions extracted by DNS and experiments has been found
 almost independent of either the flow configuration or the turbulence features. 
Its value, $D=2.23$, appears lower than that found for passive scalar isosurfaces (2.37).
The fractal scaling is lost at the inner cut-off length, $\epsilon_i$ which we observe to scale with the
dissipative Kolmogorov length, $\eta$. 
{In particular, we found that the inner cut-off length, $\epsilon_i\simeq10\eta\simeq\ell$, matches with 
$\ell\sim 10\eta$ which corresponds to the scale where the Kolmogorov's universal scaling of the 
inertial range is actually lost~\cite{sadvee}.}

The fractal features, in accordance with previous literature, e.g.~\cite{cheman,gulder1}, have been introduced in a model for 
LES of turbulent premixed flame 
in order to estimate the amount of the unresolved flame. 
Several LES have been performed and compared with reference data. A detailed comparison
between LES and corresponding DNS at $Re=6000$ demonstrated the good performance of the model in reproducing
the mean flow fields of velocity, temperature and reactants concentration.  
The LES model has been also tested at higher Reynolds numbers in order to reproduce the mean temperature
field of two experiments.  Mean flame height and width are remarkably 
estimated, reflecting a good reproduction of the mean flame evolution. 
Some differences emerged in the lower part of the shear layer and at the flame tip,
pointing out the need for the consideration of flame quenching due to 
hydrodynamic strain rates and flame stretching,  issues  which are currently
under investigation. Actually the method can be easily extended to account local quenching making null 
the wrinkling factor $\Xi$ if the local shear rate exceeds a threshold value which depends on the particular fuel mixture.
This issue is out of the scope of the present work, being still under investigation, but we aim to show
its performance in future works on this topic.   
In conclusion, the present model shows good performances in a wide range of turbulent flame regimes
allowing relatively fast simulations of premixed flames, given its simplicity (algebraic model). It 
also represents a good framework to incorporate new extensions for accounting more detailed phenomenologies such
as local quenching.

\bibliographystyle{elsarticle-num} 

\end{document}